\documentclass[12pt]{article}
\usepackage{booktabs}
\usepackage{amsmath}
\usepackage{graphicx,psfrag,epsf, xcolor}
\usepackage{enumerate}
\usepackage{natbib}
\usepackage{url} 
\usepackage[skip=2pt]{caption}
\usepackage{times,amsmath,amsfonts,amsthm,amssymb,eucal}
\usepackage{iftex}
\usepackage{xcolor}
\usepackage{algorithm}
\usepackage{algorithmic}
\usepackage{graphicx,ifthen}
\usepackage{wrapfig}
\usepackage{latexsym}
\usepackage{color}
\usepackage{mathrsfs}
\usepackage{longtable}  
\usepackage{bm}
\usepackage{bbm}
\usepackage{hyperref}
\usepackage{natbib}
\usepackage{caption}
\usepackage{tabularx}
\usepackage{float}
\usepackage{booktabs}
\usepackage{multirow}
\usepackage{xr}
\usepackage{xurl}                     
\hypersetup{hidelinks,breaklinks=true}
\newcommand{\blind}{0}

\newtheorem{theorem}{Theorem}

\newtheorem*{remark*}{Remark}

\begin{document}

\def\spacingset#1{\renewcommand{\baselinestretch}%
{#1}\small\normalsize} \spacingset{1}


\if0\blind
{
  \title{\bf Post-2024 U.S. Presidential Election Analysis of Election and Poll Data: Real-life Validation of Prediction via Small Area Estimation and Uncertainty Quantification}
  \author{Zheshi ${\rm Zheng}^{1}$, Yuanyuan ${\rm Li}^{2}$, Peter X. K. ${\rm Song}^{1}$,\\
  and Jiming ${\rm Jiang}^{3}$\\
    University of ${\rm Michigan}^{1}$, Munich ${\rm Re}^{2}$ and\\
    University of California, ${\rm Davis}^{3}$}
  \maketitle
} \fi
\date{}
\maketitle
\if1\blind
{
  \bigskip
  \bigskip
  \bigskip
  \begin{center}
    {\LARGE\bf Post-2024 U.S. Presidential Election Analysis of Election and Poll Data: Real-life Validation of Prediction via Small Area Estimation and Uncertainty Quantification}
\end{center}
  \medskip
} \fi

\bigskip
\begin{abstract}
We carry out a post-election analysis of the 2024 U.S. Presidential Election (USPE) using a prediction model derived from the Small Area Estimation (SAE) methodology. With pollster data obtained one week prior to the election day, retrospectively, our SAE-based prediction model can perfectly predict the Electoral College election results in all 44 states where polling data were available. In addition to such desirable prediction accuracy, we introduce the probability of incorrect prediction (PoIP) to rigorously analyze prediction uncertainty. Since the standard bootstrap method appears inadequate for estimating PoIP, we propose a conformal inference method that yields reliable uncertainty quantification. We further investigate potential pollster biases by the means of sensitivity analyses and conclude that swing states are particularly vulnerable to polling bias in the prediction of the 2024 USPE.
\end{abstract}

\noindent%
\vfill

\newpage
\spacingset{1.8} 
\section{Introduction}
\label{sec:intro}
On the evening of November 4, 2024, the night before the 2024 U.S. Presidential Election (USPE), there was considerable optimism in the air, at least among some of the Democrats, that their presidential candidate, Vice President Kamala Harris, would prevail in the election held on the next day, thus producing the first female President of the United States. 

There were reasons for such optimism: The polls were looking good in favor of Harris' winning odds, in particular, in most of the battle-ground states: Arizona, Georgia, Michigan, Nevada, North Carolina, Pennsylvania and Wisconsin, which showed Harris having slight leads in Michigan, Pennsylvania and Wisconsin, while virtually tied with her opponent, former President Donald Trump, in Georgia and North Carolina. The electoral-college (EC) calculation showed that Harris did not need to win all such states, or even the majority of these battle-ground states. In fact, if she could just ``cash in'' the lead she had in the polls of Michigan, Pennsylvania and Wisconsin, she would be all but guaranteed to become the first female U.S. President. To add further optimism, the latest polls seemed to be showing Harris even taking a lead in some of the Republican strongholds, such as Iowa. In contrast, Trump would have to win almost every battle ground state in order to win, not to mention losing any of his strongholds---his road to return to the White House appeared to be much tougher.

What about the polls' under-prediction of Trump's support among the voters, a well-known fact from the 2016 and 2020 elections? Well, the Democrats believed, as did most, if not all, of the election pollsters, that such a potential under-prediction bias had been ``factored'' into the polling results. In fact, the polls did do better in predicting the 2020 election, as compared to the 2016 election, and after 2020 there were four more years in the making to improve the poll's prediction accuracy. Finally after eight long years, on the USPE eve (i.e., night before the USPE), someones were optimistically looking forward to seeing the polls, well, do what the polls do, that is, accurately predicting the USPE results. Really?

Kamala Harris ended up losing all of the seven battle-ground states. As a result, Donald Trump has made the most astonishing, and mathematically ``improbable'', political comeback in the U.S. history. In particular, he easily held the Republican stronghold Iowa, by a large margin, making the latest poll result look embarrassingly misleading.

This motivated many statisticians, including us, to curiously investigate possible causes for this landsliding loss to Harris. As noted, after the disastrous failure of the polls in 2016, the pollsters were seemingly making progress, and did better in 2020. But then came this, the 2024 USPE. One could argue that, at the very least, the polls had correctly predicted Hillary Clinton winning the popularity vote in 2016, something they did not even get right in 2024. This brings back, once again, the same old question: Why is it so hard to predict a Trump election? We call it a Trump election because all three elections, 2016, 2020, and 2024, involved the same Republican candidate, Donald Trump.

While the pollsters, and political scientists, may be looking for answers, the answers are expected to be anything but simple, and possibly not to be found for many years, if at all. On the other hand, there are methods, including statistical methods, that demonstrably better predict election results, especially for the USPE; see, for example, \cite{ferrjohn1974}, \cite{merrill1978}, \cite{gelman1993}, \cite{gelman1994}, \cite{rusk2001}, \cite{katz2002}, FiveThirtyEight (https://fivethirtyeight.com/), and \cite{jls2023}. This may be related to what the pollsters called were ``factored in'' in the 2024 polls. Yet, it matters what were factored in, and how to factor them in.

Interestingly, a statistical method, proposed by \cite{jls2023}, performs much better at correctly predicting the 2024 USPE results. Their prediction machinery is built upon the theory of small area estimation (SAE; e.g., \cite{raomolina2015}). Out of the 50 states of the U.S. plus the District of Columbia (D.C.), seven of them, namely, Alabama, D.C., Hawaii, Idaho, Kentucky, Louisiana, and Mississippi, did not have any polls conducted after Harris replaced Joe Biden as the Democratic candidate on August 5, 2024; see, for example, \cite{APNews2024}. For the remaining 44 states of the U.S., the SAE-based prediction (SAEP) method of \cite{jls2023} correctly predicts the winner, and therefore the EC result, for every single one of those 44 states, leading to a total of 219 EC votes for Harris vs 277 EC votes for Trump, which is exactly the same outcomes seen in the real election. In contrast, the benchmark method based on average of the polls incorrectly predicted the results in the battleground states of Michigan, Pennsylvania, and Wisconsin, hence yielding a (very) wrong total EC votes, 263 for Harris vs 233 for Trump, which would have changed the outcome of the 2024 USPE. We refer to more detailed results in Section \ref{sec: sae prediction}. Notably, the well-known political website, 538 (https://fivethirtyeight.com/) among others, also made incorrect predictions using the polling average method.

Although the empirical results suggest that the SAEP achieves remarkably high accuracy, a critical question remains: How confident are we about the prediction results before we know the truth? Also, because the predictions would be made one week before the actual election, unforeseen events during that period may affect the results. It is therefore desirable to quantify the probability of incorrect prediction (PoIP) for the projected EC winner in each state. Adding this additional rigor to the prediction matters as it addresses uncertainty beyond the polling data. This motivates us to invoke and extend conformal prediction to estimate the PoIP for each state-level prediction. We also perform a sensitivity analysis to demonstrate the impact of pollster-specific biases on the reliability of the predictions and distort the associated uncertainty measures.

These are the main lines of stories in the paper. In Section 2, we first provide some preliminaries for USPE and introduce the notations to be used subsequently. In Section 3, we describe the SAEP method under the notion of transfer learning, and report the detailed prediction results for the 2024 USPE. The conformal uncertainty quantification method with a sensitivity analysis is discussed in Section 4. Additional analysis results are presented, and discussion and concluding remarks are offered in Section 5.
\section{Preliminaries}
\label{sec:prelim}

\subsection{The EC system of USPE}
\label{sec:upse} 

Here is a brief summary of the EC system for the USPE.  The EC assigns each state a certain number of EC votes, with larger states (in terms of the population) receiving more EC votes. If a candidate wins a state by simple majority, the candidate collects all the EC votes attached to the state; this applies to all except two states, Maine and Nebraska. For the latter two states, the state is divided into districts with each district holding one EC vote; different districts can be won by different candidates so, as a result, the state's total EC votes may be split between the two candidates. The numbers of EC votes then add up over the 50 states and D.C. to yield the total of EC votes for each candidate, and whoever receives 270 or more EC votes wins the presidency.

\subsection{Polling-based models for USPE prediction}\label{sec: notation}

Polling-based models have become fundamental for the prediction of USPE results, moving beyond simple aggregation to complex statistical methodologies accounting for political heterogeneity and uncertainty in this sophisticated country, and election system, too. These models, exemplified by platforms like \cite{fivethirtyeight_polling_averages}, aim primarily to correct for potential biases in individual polls by weighting them in terms of factors such as pollster quality, sample size, and recency. This weighting approach is highly responsive to the dynamics of a campaign but can be susceptible to systematic polling errors, as seen in recent USPEs \citep{Durand2023,Barnett2023}. Furthermore, the winner-take-all mechanism of the EC system, which necessitates accurate state-level predictions, poses a significant challenge to yield a prediction of high accuracy. Polling data for specific ``swing states'' can be sparse, inconsistent or noisy, making precise predictions notoriously difficult \citep{Gelman2021}.

To address the challenge of limited substate data, researchers have increasingly adopted small area estimation (SAE) methods to estimate state opinion and political propensity, see \cite{lax2009, dejonge2018}. These methods ``borrow strengths" from other geographic areas and temporal time windows to generate more robust and reliable estimates of political leaning for small geographic areas, demographic groups, or geographic locations crossed by demographic groups, where direct survey data are often insufficient and potentially unbalanced (e.g., \cite{raomolina2015}).

A recent development in this field, particularly for SAE-based prediction (SAEP) models, pertains to the invocation of transfer learning to leverage data from previous elections. \cite{jls2023} adopted this technique to train an SAEP model with the election and poll data of the 2016 USPE, which was then applied to predict the 2020 USPE using the poll data of the 2020 USPE. The underlying assumption is that there is a legitimate linkage between these two USPEs, pivotal to the consistent presence of the same Republican candidate. Their SAEP approach takes advantage of the consistent voting patterns and demographic shifts over time, which can be particularly useful when current data are limited. By transferring knowledge from a prior, well-learned election, the resulting SAEP model can make more informed and robust predictions, as long as the core political dynamics remain relevant, which is arguably the case in the 2020 USPE and even more so in the 2024 USPE.

\subsection{Notation}
One of the primary objectives in this paper is to validate the SAEP model proposed by \cite{jls2023} using independent external data from the 2024 election and related polls. This validation is surely interesting as their model had been developed prior to the 2024 USPE, so that no knowledge, what-so-ever, about the 2024 election would have been available to make any possible calibration of the model.  To proceed, we introduce some relevant notation.

For both 2016 and 2020 USPEs that are used as training data for the SAEP model, let $p_{ijk}$ denote the final-week poll result (expressed as a proportion of vote share) in state $i$ reported by pollster $j$ for party $k$, where $i = 1, 2, \ldots, 51$, $j =1,2,\cdots, n_i$, while index $k = 1, 2$ corresponds to the Democratic and Republican candidates, respectively. Similarly, let $\pi_{ik}$ denote the actual realized election result recorded as a proportion ({\it a.k.a.} support rate) of votes supporting party $k$ in state $i$. We define the ``Democratic over Republican (DoR)" margin, $d_i = \log(\pi_{i1}/\pi_{i2})$, by the log‑odds of the support for the Democratic candidate over the Republican candidate in state $i$. Clearly, the Democratic candidate wins in state $i$ if its DoR margin $d_i > 0$, while the Republican candidate wins if the DoR margin $d_i < 0$. Ties are technically nearly impossible,  and indeed not observed in our data, so they would be therefore excluded from the analysis if they occurred.

For our prediction target of the 2024 USPE, we use a superscript $*$ to distinguish quantities related to this year. Specifically, let $p_{ijk}^*$ denote the final-week poll result for state $i = 1, 2, \ldots, 44$, pollster $j =1,2,\cdots,n_i^*$, and party $k = 1, 2$. Let $\pi_{ik}^*$ represent the actual corresponding election result, or support rate.  The realized DoR margin is denoted by $d_i^* = \log(\pi_{i1}^* /\pi_{i2}^*)$.

As usual in the statistical literature, we use a hat $\hat{\cdot}$ to denote estimated or predicted quantities. For instance, we denote the small area estimate for a certain polling bias by $\hat{\theta}_{ik}$ and a predicted support rate in 2024 by $\hat{\pi}_{ik}^*$. Moreover, we denote a predicted DoR margin as $\hat{d}_i^* = \log(\hat{\pi}_{i1}^* / \hat{\pi}_{i2}^*)$.

\section{Prediction of 2024 USPE}\label{sec: sae prediction}
\label{sec:back}
\subsection{Transfer-learning-based prediction via SAE} 
We are interested in examining the SAEP method introduced in \cite{jls2023} for its performance in predicting the 2024 USPE. 
The outcome, or response, of the SAEP model is $y_{ijk} = \log(p_{ijk}/\pi_{ik} ) = \log(p_{ijk} )-\log(\pi_{ik} )$, characterizing the polling bias. For example, $y_{ijk} < 0$ means that the poll rate by pollster $j$ underestimates the true support rate for party $k$ in state $i$. We focus on the two SAEP models (i.e. Model I and Model III) given in \cite{jls2023}, which take the following forms:
\begin{eqnarray}
&&{\rm Model}\;{\rm I:}\;\;\;y_{ia}=\beta_{0}+\beta_{1}I_{{\rm can},ia}+z_{ia}v_{i}+e_{ia},
\\
&&{\rm Model}\;{\rm III:}\;\;\;y_{ijk}=\beta_{0}+\beta_{1}I_{{\rm can},ijk}+z_{ijk}v_{i}+u_{j}
+e_{ijk}, \label{model 3}
\end{eqnarray}
where $a = 1,2,\cdots, n_i$,
$n_i$ denotes the
total number of combined indices (i.e. the sample size) for state $i$. Furthermore, $I_{can,ia}$ and $I_{can,ijk}$ are dummy
variables for candidates,  which equal to 0 for Democrat, and 1 for Republican; $z_{ia} = (1, 0)$ for Democratic, and $z_{ia} = (0, 1)$ for Republican and so is $z_{ijk}$. Moreover, $v_i\sim N(0, G)$ denotes the 2-dimensional state-level random effect with covariance matrix $G = \begin{pmatrix}
\sigma_d^2 & \rho \sigma_d \sigma_r \\
\rho \sigma_d \sigma_r & \sigma_r^2
\end{pmatrix}$, while $u_j\sim  N(0,\sigma^2)$ corresponds to  the pollster level random effect. Following \cite{jls2023}, we assume that the errors $e_{ia}$ and $e_{ijk}$ are independent and follow the $N(0,\tau^2)$ distributions. All the variance components in the models,  including $\tau$, $G$, and $\sigma$ are unknown.

Model I and Model III are trained by data consisting of the election and poll results from both 2016 and 2020 USPEs, where the model parameters are estimated by the restricted maximum likelihood (REML) method implemented by the R package {\it lme4}. The estimated parameters are presented in Table \ref{tab:REML Model I & III}, in which the standard errors of the variance components estimates were obtained via the bootstrap method. All fixed effects (i.e., $\beta$ parameters) are significant at the 5\% significance level. In both models, the estimate of $\hat{\beta}_1$ is negative, indicating that the overall nation-level poll underestimates the support rate of the Republican candidates. This finding is in agreement with that reported in \cite{jls2023}, where various similar SAEP models are trained based on either 2016 or 2020 USPE data. The variance estimators are also significant, suggesting reliability of the SAEP models to capture key signals and features from such data.

A major advantage of the SAEP method is that it provides not only estimation but also prediction of mixed effects through fixed effects and area-specific random effects, which are deemed unique strengths offered by the small-area estimation methodology. By plugging in these estimates of the training models, we obtain the empirical best linear unbiased prediction (EBLUPs) of the small-area mean, $\theta_{ik}$, which corresponds to the average polling bias in state $i$ for party $k$ and pollster $j$,
\begin{align}
&{\rm Model}\;{\rm I:}\;\;\;\hat{\theta}_{ijk}=\hat{\beta}_{0}+\hat{\beta}_{1}1_{(k=2)}+\hat{v}_{ik},\label{eq: EBLUP of Model I}\\
&{\rm Model}\;{\rm III:}\;\;\;\hat{\theta}_{ijk}=\hat{\beta}_{0}+\hat{\beta}_{1}1_{(k=2)}+\hat{v}_{ik}+\hat{u}_j,\label{eq: EBLUP of Model III}
\end{align}

Similarly, for the 2024 USPE, the SAEP model can be expressed as follows:
\begin{eqnarray}
\log(p_{ijk}^{*})-\log(\pi_{ik}^{*})&=&\theta_{ijk}^{*}+e_{ijk}^{*},\label{sa_mean_mdIj}
\end{eqnarray}
where $e_{ijk}^{*}$ corresponds to the random error term $e_{ijk}$ in (\ref{model 3}).
Averaging both sides of (\ref{sa_mean_mdIj}) over $j=1,2,\ldots,n_i^*$, we get
\begin{eqnarray}
\overline{\log(p_{i\cdot k}^{*})}-\log(\pi_{ik}^{*})&=&\bar{\theta}_{i\cdot k}^{*}+\bar{e}_{i\cdot k}^{*},\label{sa_mean_mdI}
\end{eqnarray}
where $\overline{\log(p_{i\cdot k}^{*})}=(n_i^*)^{-1}\sum_{j=1}^{n_i^*}\log(p_{ijk}^{*})$, $\bar{\theta}_{i\cdot k}^{*}=(n_i^*)^{-1}\sum_{j=1}^{n_i^*}\theta_{ijk}^{*}$ and
$\bar{e}_{i\cdot k}^{*}=(n_i^*)^{-1}\sum_{j=1}^{n_i^*}e_{ijk}^{*}$. If we replace $\theta_{ijk}^{*}$ by the $\hat{\theta}_{ijk}$ in (\ref{eq: EBLUP of Model I}) or (\ref{eq: EBLUP of Model III}), and ignore the second term in the right side of (\ref{sa_mean_mdI}) which is expected to be negligible, we obtain
\begin{eqnarray}
\pi_{ik}^{*}\approx\hat{\pi}_{ik}^{*}\equiv\exp\left\{\overline{\log(p_{i\cdot k}^{*})}-\hat{\theta}_{ik}\right\},
\;\;\;i=1,\dots,51,\;\;k=1,2,\label{sae_pred_ik}
\end{eqnarray}
where $\hat{\theta}_{ik}=(n_i)^{-1}\sum_{j=1}^{n_i^*}\hat{\theta}_{ijk}$. Of note, the omission of the average error term $\bar{e}_{i\cdot k}^{*}$ is supported by the law of large numbers, as this is an average of a large number of mean-zero random errors. 

\subsection{Prediction results of 2024 USPE}\label{sec: 2024 predction}

\begin{figure}
    \centering
    \includegraphics[width=0.9\linewidth]{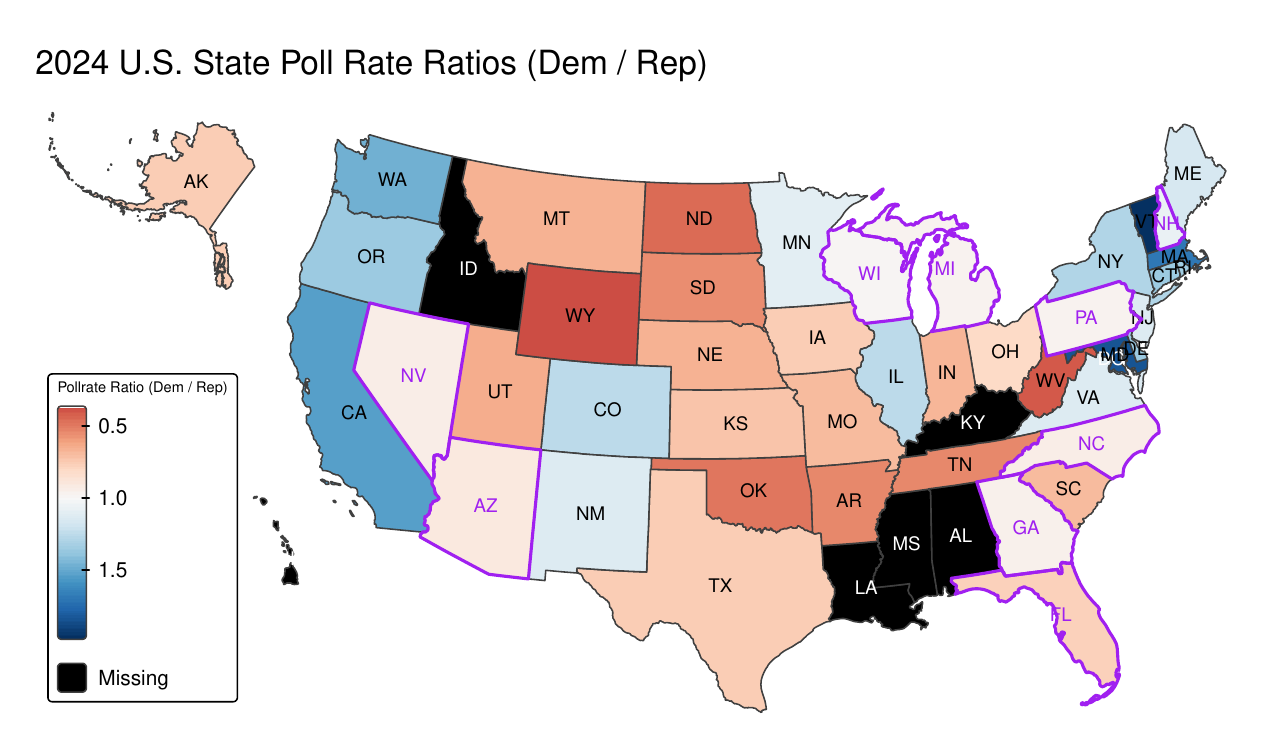}
    \caption{2024 U.S. presidential election results. States shaded in darker blue indicate stronger Democratic victories (i.e., larger values of $\exp(d_i^*)$), while darker red indicates stronger Republican victories (i.e., smaller values of $\exp(d_i^*)$). States outlined in purple represent swing states.}
    \label{fig: us map}
\end{figure}

We now apply (\ref{sae_pred_ik}) to predict the $\pi_{ik}^{*}$ for the 2024 USPE. Similar to \cite{jls2023}, we download all eligible polls data for the 2024 presidential election from the 538 website: \url{https://projects.fivethirtyeight.com/2024-election-forecast/}. 
The dataset includes polls from 44 states, that is, the 50 states and D.C. with seven states (Alabama, District of Columbia, Hawaii, Idaho, Kentucky, Louisiana, and Mississippi) excluded due to the absence of polls after August 5, 2024 when Kamala Harris replaced Joe Biden as the Democratic nominee. Given the historically stable voting patterns in these seven blue or red states, we use their 2020 election outcomes as proxies for prediction for the technical needs but exclude them from results reporting.
As of now, the actual confirmed 2024 election results are available. We display the actual election results in Figure~\ref{fig: us map}. Among these 44 states with polling data, we label nine swing (purple) states that are outlined in purple in Figure~\ref{fig: us map}) because of particular interest in further analyses; they are, Nevada, Arizona, Wisconsin, Michigan, Pennsylvania, North Carolina, New Hampshire, Georgia, and Florida. By a swing state we mean its election was flipped from one party to the other during the past three presidential elections (2012, 2016, and 2020) \citep{USNews2024}, or had consistently narrow margins in both 2016 and 2020 elections \citep{USAFacts2025}. These are deemed the battleground states that ultimately determined the USPE outcome. 
In the remaining part of the paper, we treat the officially confirmed 2024 results as the oracle targets in the evaluation of prediction performances based on the polling data collected one week before the 2024 election.

We train the above SAEP models on the combined dataset consisting of both the 2016 and 2020 election results and corresponding poll data, and predict the 2024 USPE outcomes using the SAEP adjusted poll predictions, $\hat{\pi}_{ik}^{*}$, for the 44 states. We then compare the prediction results with the actual outcomes of the 2024 USPE. We also compare the SAEP method with the ``poll of polls" (PoP), i.e., a simple average of polls results to get final support rate prediction, given by
\begin{eqnarray}
\bar{p}_{i\cdot k}^{*}&=&\frac{1}{n_i^*}\sum_{j=1}^{n_i^*}p_{ijk}^{*},\;\;\;i=1,\dots,51,\;\;k=1,2.\label{pop_pred_ik}
\end{eqnarray}
The predicted winners using either our SAEP models or the PoP method are presented in Table \ref{tab: short prediction results}, in which we intentionally omit the states for which all methods have correctly predicted the winners. Interestingly, Model I and Model III have reached the same prediction outcomes in all states that perfectly match the actual election results, yielding a 100\% accuracy. The PoP method incorrectly predicted the outcomes of the three important battleground states, Michigan, Pennsylvania and Wisconsin, leading to an incorrect prediction of the final national winner.

\begin{figure}[hbtp!]
    \centering
    \includegraphics[width=0.9\linewidth]{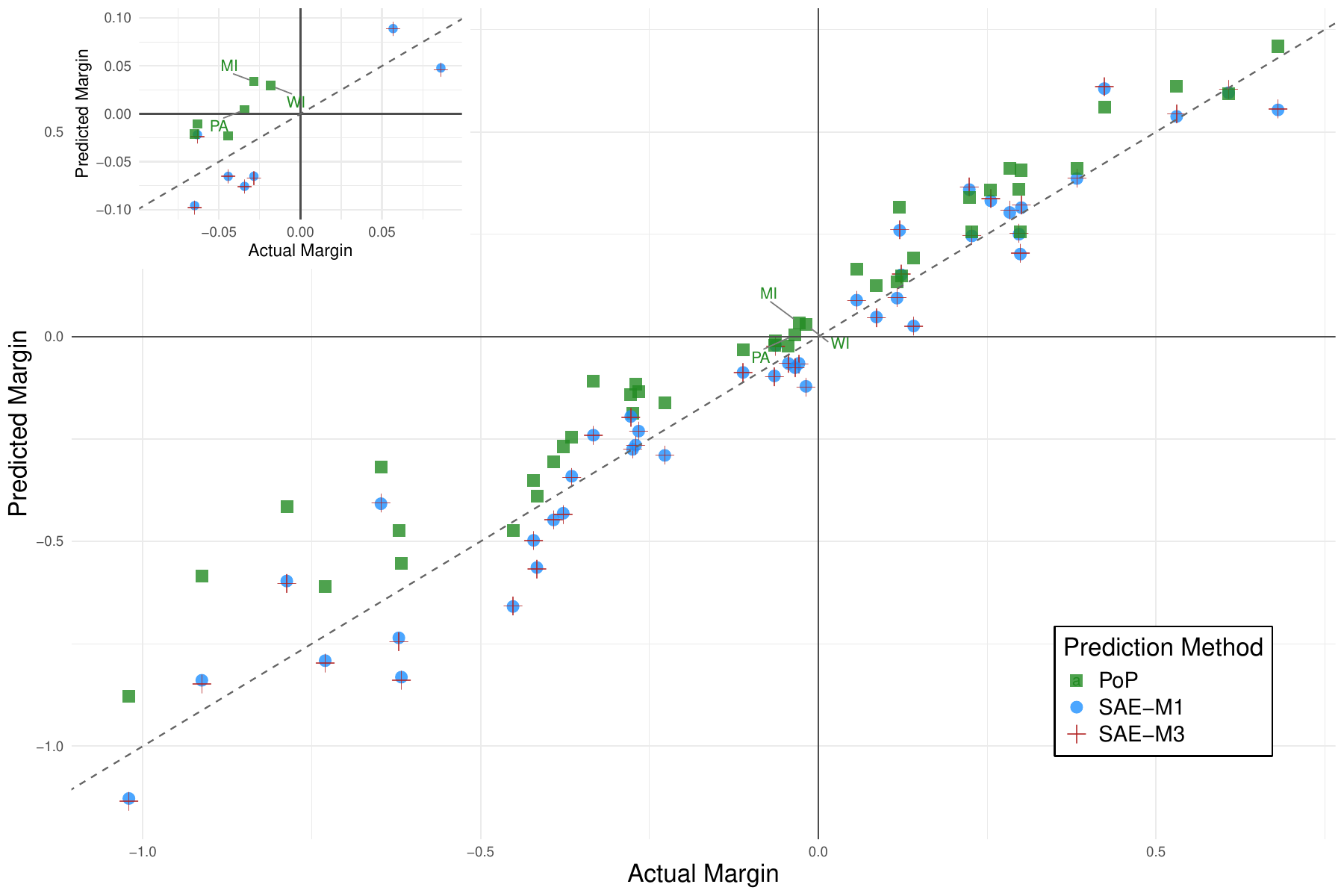}
    \caption{Predicted DoR margins $\hat{d}_i^*$ vs. Actual DoR margins $d_i^*$ for 2024 USPE in the 44 states with valid polls data. The points' distances to the $y=x$ dash line show deviation of the predictions. Points in Quadrant II and IV are incorrect winner predictions, which are PoP predictions for the states MI, WI and PA. A zoomed-in plot around the origin is shown in the top-left corner of the figure.}
    \label{fig:predicted_margin}
\end{figure}

To evaluate the prediction accuracy of support rate, we draw the predicted DoR margin $\hat{d}_i^*$ over $d_i^*$ in a scatterplot shown in Figure \ref{fig:predicted_margin}. The graph reveals that the PoP method demonstrates a clear overestimation of the support rate for the Democratic candidate, judged by its predicted margins consistently exceeding the actual USPE results. In contrast, the predicted DoR margins from two ASEP Models I and III are tightly distributed around the $y = x$ diagonal line of perfect prediction, indicating high prediction accuracy with no obvious partisan biases. Overall, the two SAEP models perform very similarly and exhibit slightly larger errors for strong Republican states. In contrast, the PoP method shows three critical errorous predictions highlighted in Quadrant II for Michigan, Pennsylvania, and Wisconsin. In these three states, the PoP method predicted a Democratic victory (i.e. a positive margin) when the actual result was a Republican win (i.e. a negative margin). The detailed prediction results for the 44 states are presented in Table S1 of the Supplementary Material.

Since the SAEP Model I and Model III yield similar prediction results for all of the $44$ states, for parsimony consideration in the model building, we adopt the simpler Model I for further discussion for uncertainty quantification in Section~\ref{sec:uq}. 

\begin{table}[hbtp!]
\begin{center}
\vspace{2mm}
{\footnotesize
\begin{tabular}{cccccc}
Model & Parameter        & \multicolumn{1}{c}{Estimate} & \multicolumn{1}{c}{Standard Error} & \multicolumn{1}{c}{t-statistic} & Group      \\ \hline
I     & $\beta_{0}$      & -0.023   & 0.009	 & -2.628    & Fixed      \\
      & $\beta_{1}$      & -0.088	 &0.019	 &-4.593  & Fixed      \\
      & $\sigma_{\rm d}$ & 0.056    & 0.007          & 8.598     & State      \\
      & $\rho$           & -0.826   & 0.074         & -11.200   & State      \\
      & $\sigma_{\rm r}$ & 0.081    & 0.009          & 8.240   & State      \\
      & $\tau$           & 0.088    & 0.002         & 52.100    & Residual   \\ \hline
III     & $\beta_{0}$      & -0.025	&0.009	 &-2.645    & Fixed      \\
      & $\beta_{1}$      & -0.086	&0.020	&-4.322    & Fixed      \\
      & $\sigma$         & 0.037    & 0.004          & 10.259    & Pollster   \\
      & $\sigma_{\rm d}$ & 0.057    & 0.007          & 8.132     & State      \\
      & $\rho$           & -0.733   & 0.081          & -9.097    & State      \\
      & $\sigma_{\rm r}$ & 0.092    & 0.011          & 8.504     & State      \\
      & $\tau$           & 0.070   & 0.001          & 54.783    & Residual   \\ \hline
\end{tabular}}
\caption{REML estimates for Model I and III}
\label{tab:REML Model I & III}
\end{center}
\end{table}

\begin{table}[hbtp!]
\begin{center}
\vspace{2mm}
{\footnotesize
\begin{tabular}{llllr}
State&Actual&SAEP&PoP&EC Votes\\ \hline
Michigan &R & R & D &15\\
Pennsylvania &R &R & D &19\\
Wisconsin &R &R &D  &10\\ \hline
National&R (226/312)&R (226/312)&D (270/268)
\end{tabular}}
\caption{Comparison of prediction results of different methods for 2024 USPE. Only states where at least one method yields incorrect prediction are presented. The total EC counts include empirical prediction based on the U.S. political map of the 7 states without the polls after the Biden-Harris changeover, which is the same for both SAEP and PoP.}
\label{tab: short prediction results}
\end{center}
\end{table}

\section{Prediction uncertainty quantification}
\label{sec:uq} 
In practice, it is essential not only to generate a point prediction but also to assess the confidence on the given prediction subject to data sampling variability. Arguably,  uncertainty quantification is deemed a critical component attached with the prediction analysis. In our setting, the prediction target is a categorical outcome labeling the winning candidate in a given state. Conventional uncertainty measures, such as prediction intervals for $\pi_{ik}$, are not readily applicable here because values outside the interval may still correspond to the same election outcome. To address this issue, we propose using the probability of incorrect prediction (PoIP), whose formal definition is given below, as a measure of uncertainty.

Given the actual and predicted DoR margins $ d_i^* = \log(\pi_{i1}^* / \pi_{i2}^*)$ and  $\hat d_i^* = \log(\hat \pi_{i1}^* / \hat \pi_{i2}^*)$ (as defined in Section~\ref{sec: notation}), we formally define the PoIP for state $i$ as follows:
\begin{eqnarray*}
   {\rm PoIP}_i = \left\{\begin{array}{cc}P(\hat d_i^*>0\mid d_i^*<0),&{\rm if}\; d_i^*<0;\\
P(\hat d_i^*<0\mid d_i^*>0),&{\rm if}\; d_i^*>0.\end{array}\right.
\end{eqnarray*}
A large absolute value $|d_i|$ indicates a strong partisan lead in state $i$, while a small absolute value $|d_i|$ suggests that the state is likely to be a swing state. In practice, we are particularly interested in evaluating the PoIP for swing states, where prediction uncertainty plays a more critical role in understanding and interpreting a predicted outcome. 

\subsection{Prime of conformal prediction}

A common approach to estimating the PoIP is via the bootstrap method, which approximates the standard deviation of the distribution of predictions, which is then used to approximate the PoIP. However, our preliminary analysis shows that the bootstrap-based approach systematically underestimates the standard deviation (see \autoref{boostrap PoIP}), leading to implausibly small PoIP values. For example, even in swing states such as Michigan and Pennsylvania, the bootstrap method yields PoIP values below $10^{-5}$. Such minimal uncertainty is obviously not aligned with the real-world political leaning behaviors associated with these states. This motivates the use of an alternative approach based on conformal prediction.

Conformal prediction, originally introduced by \citet{shafer2008tutorial,vovk2005algorithmic}, attempts to construct prediction intervals that remain valid under model misspecification by accounting for both bias and variance \citep{xie2022homeostasis}. Empirical evidence indicates that conformal methods generally produce wider intervals when the prediction model is biased, thereby ensuring valid coverage. In our application, this robustness allows conformal prediction to correct the downward bias in variance estimation inherent in the bootstrap method. Furthermore, we investigate the impact of pollster bias on conformal-based PoIP using sensitivity analysis. We find that pollster bias can lead to overestimation of PoIP in battleground states but underestimation in strongly partisan (blue or red) states, offering practical insights for the interpretation and application of conformal-based PoIP in the USPE prediction. See more details of the conformal method in Section~\ref{sec:conformal}.

\subsection{Bootstrap estimation of PoIP}\label{boostrap PoIP}
We begin with the following inequality for a given value $d\in[-1,1]$:
\begin{eqnarray}
P\big(\hat{d}_i^*> d|d_i^*<0)\big)&=&P\left[\left.\frac{\hat d_i^*-d_i^*}{{\rm sd}(\hat{d}_i^*)}>\frac{d-d_i^*}{{\rm sd}(\hat{d}_i^*)}\right|d_i^*<0\right]\nonumber\\
&\leq&P\left[\left.\frac{\hat{d}_i^*-d_i^*}{{\rm sd}(\hat{d}_i^*)}>\frac{d}{{\rm sd}(\hat{d}_i^*)}\right|d_i^*<0\right]. \label{eq: poip_1}
\end{eqnarray}
The last inequality of Equation~(\ref{eq: poip_1}) suggests that the approximation becomes tighter for smaller values of $|d_i^*|$, corresponding to plausible swing states that are really influential states in USPE.

If we assume 
\[
\frac{\hat{d}_i^*-d_i^*}{{\rm sd}(\hat{d}_i^*)}\stackrel{\cdot}{\sim}N(0,1),  
\]
where $\stackrel{\cdot}{\sim}$ denotes “approximately distributed as”, then by letting $z_i(d) = d / {\rm sd}(\hat{d}_i^*)$ we obtain
\begin{equation*}
{P}(\hat{d}_i^*> d|d_i^*<0)\le 1-\Phi\big(z_{i}(d)\big).
\end{equation*}
Here $\Phi(\cdot)$ denotes the cumulative distribution function (CDF) of the standard normal distribution. We then need to estimate the standard deviation of the predicted margin, $sd(\hat{d}_i^*)$. The standard bootstrap method for estimating $sd(\hat d_i^*)$, commonly used in the SAE literature (e.g., \cite{raomolina2015}), is described in Section S1 of the Supplementary Material. Replacing ${\rm sd}(\hat{d}_i^*)$ with its bootstrap estimate ${\rm se}_{boot}(\hat{d}_i^*)$ yields $\hat{z}_{i}(d) = d/{\rm se}_{boot}(\hat d_i^*)$, and consequently,
\begin{equation*}
{P}(\hat{d}_i^*> d|d_i^*<0)\le 1-\Phi\big(\hat{z}_{i}(d)\big).\label{poip_3}
\end{equation*}

Similarly, we obtain the inequality for the case where $d_i^*>0$:
\[
P(\hat d_i^*<d\mid d_i^*>0)\le \Phi(\hat z_i(d))
\]
When $d$ is the value of $\hat d_i^*$ from our SAEP model, we plug it into $\hat z_i(d)$ 
This leads to the following approximate upper bounds for PoIP:
\begin{eqnarray}
   {\rm PoIP}_i \le  \left\{\begin{array}{cc}1-\Phi\big(\hat{z}_i(\hat d_i^*)\big),&{\rm if}\; d_i^*<0;\\
\Phi\big(\hat{z}_i(\hat d_i^*)\big),&{\rm if}\; d_i^*>0.\end{array}\right.\label{poip_5}
\end{eqnarray}
We refer the right side of (\ref{poip_5}) to as the bootstrap estimator of PoIP. We present the results of those bootstrap PoIPs in Table~\ref{tab: PoIP boot selected}, including the nine swing states and two non-swing states (Nevada and Maine) selected due to their larger $\rm PoIP>0.01$ than all the other states.  The full results for all states are presented in Table S2 in the Supplementary Material.

We observe from Table~\ref{tab: PoIP boot selected} that the bootstrap PoIP estimates are too small for all of the battleground states, indicating overly high confidence on predictions for those states. The bootstrap PoIP in Table~\ref{tab: PoIP boot selected}  are arguably misleading and incorrect, due primarily  to the fact that the bootstrap estimate of $sd(\hat d_i^*)$ appears to be too small to capture realistically the substantial uncertainty in these states for the 2024 USPE. One possible explanation is that the conventionally bootstrap resampling approach for binary outcomes may be non-informative and insensitive to the underlying uncertainty, resulting in severely underestimated variances.

The need to improve the uncertainty quantification in the analysis has motivated us to seek for an alternative approach to yield more reliable and trustworthy PoIP estimates. In the next section, we investigate a popular uncertainty quantification technique called conformal prediction to overcome this underestimation issue and provide more robust and reliable uncertainty measures.
\begin{table}[htbp]
\centering{\footnotesize
\begin{tabular}{lrrrccc}
\hline
\textbf{State} & \textbf{real-rate OR} & \textbf{SAE OR} & \textbf{PoP OR} & \textbf{bootstrap se} & \textbf{bootstrap PoIP}&\textbf{conformal PoIP} \\
\hline
FL & 0.766 & 0.795 & 0.876 & 1.010 & $<0.0001$ & 0.047 \\
OH & 0.797 & 0.749 & 0.851 & 1.013 & $<0.0001$ & 0.035 \\
AZ & 0.894 & 0.918 & 0.969 & 1.011 & $<0.0001$ & 0.494 \\
GA & 0.957 & 0.938 & 0.977 & 1.011 & $<0.0001$ & 0.494 \\
NC & 0.937 & 0.909 & 0.980 & 1.010 & $<0.0001$ & 0.494 \\
NV & 0.939 & 0.978 & 0.991 & 1.013 & 0.038    & 0.494 \\
PA & 0.966 & 0.927 & 1.005 & 1.010 & $<0.0001$ & 0.494 \\
MI & 0.971 & 0.936 & 1.035 & 1.010 & $<0.0001$ & 0.494 \\
WI & 0.983 & 0.885 & 1.030 & 1.011 & $<0.0001$ & 0.335 \\
NH & 1.058 & 1.093 & 1.179 & 1.017 & $<0.0001$ & 0.029 \\
ME & 1.153 & 1.027 & 1.211 & 1.022 & 0.113    & 0.064 \\
\hline
\end{tabular}}
\caption{Estimated PoIP from both bootstrap and conformal approaches (bootstrap PoIP and conformal PoIP) for nine selected battleground states or two states with bootstrap PoIP $>0.01$. Reported quantities include the odds ratios of the real-rate (real-rate OR), the SAE-estimated odds ratios (SAE OR), the PoP odds ratios (PoP OR), and the bootstrap-estimated standard error of the odds ratio (bootstrap se). States are ordered by the odds ratios of the real-rate.}
\label{tab: PoIP boot selected}
\end{table}

\subsection{Conformal prediction method for estimating PoIP}
\label{sec:conformal}
Let the observed data be the 2016 and 2020 election and posters' polling data and denote them as $\mathcal{D}_{\rm obs}$. Denote our prediction target of 2024 election and posters' polling data as $\mathcal{D}_{\rm pred}$. For any dataset $\mathcal{D}_{\rm index}$, let $\mathcal{I}_{\rm index}$ denote its associated index set.
To ensure computational efficiency, we adopt the split conformal procedure \citep{lei2018distribution}, and the detailed introduction of the standard split conformal method is given in the Section S2 in the Supplementary Material.
Our strategy of data split $\mathcal{D}_{obs} = \mathcal{D}_{tr}\cup\mathcal{D}_{ca}$ goes along with the election year, where $\mathcal{D}_{tr}$ and $\mathcal{D}_{ca}$ denote the training and calibration datasets. The most important assumption of conformal prediction is the exchangeability between the calibration and the prediction target (i.e $\mathcal{D}_{\rm ca} \text{ and }\mathcal{D}_{\rm pred}$ are drawn from the same distribution and their joint distribution is invariant under permutations of the sample order.). Notably, the 2016 election shares more structural similarities with the 2024 election: in both years, the Republican candidate is Donald Trump; the incumbent administration is Democratic; and, in hindsight, the media and polling data leading up to the election were overly confident in a Democratic victory. These parallels give a defensive argument for the exchangeability assumption to hold between 2016 and 2024. Therefore, we choose the 2016 election outcomes data as the calibration set $\mathcal{D}_{\rm ca}$ and the 2020 election outcomes data as the training set $\mathcal{D}_{\rm tr}$. For completeness, we also conduct another analysis reversing the 2020 election data as $\mathcal{D}_{\rm ca}$ and the 2016 data as $\mathcal{D}_{\rm tr}$. The corresponding results in the second analysis are reported in Figure S1 of the Supplementary Material.

\begin{figure}[h!tbp]
    \centering
    \includegraphics[width=0.8\linewidth]{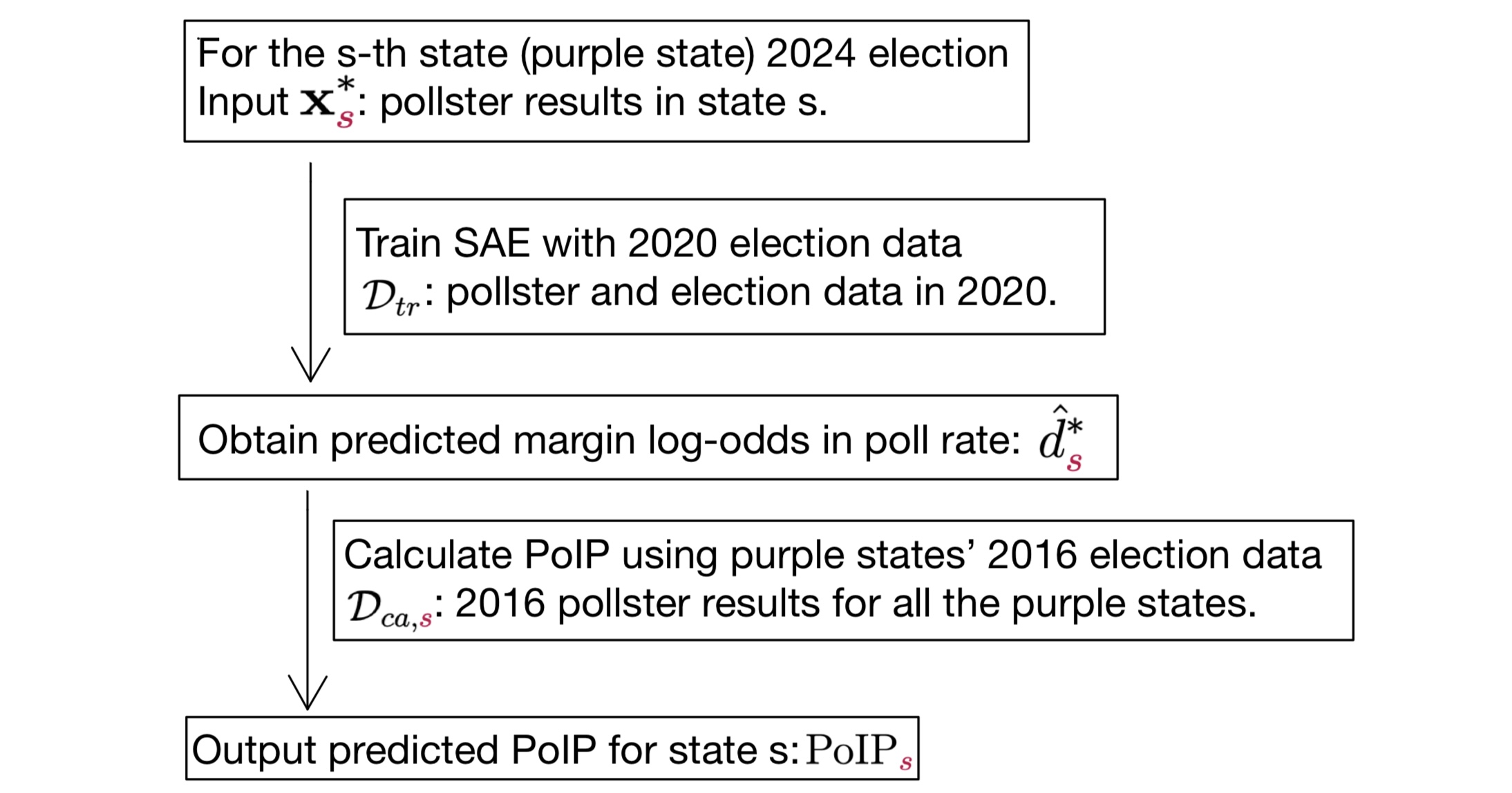}
    \caption{Workflow diagram of calibration data enrichment for estimating PoIP using conformal prediction in a purple (swing) state $s \in \mathcal{I}_{\rm purple}$. The same procedure applies to red and blue states with appropriate calibration sets.}
    \label{fig: workflow CP PoIP}
\end{figure}

For the prediction of state $s$ in the 2024 election, we account for differences in the distribution of polling and election data across political leanings by constructing a localized calibration set. Specifically, we include only the 2016 election data from states that share the same political leaning as state $s$, and denote this localized calibration set as $\mathcal{D}_{\text{ca},s}$. This strategy, which helps ensure the exchangeability between the calibration set and the prediction target, follows the idea proposed in \cite{guan2023localized}.
To address this, we adopt a data augmentation strategy that leverages the availability of pollster-level poll results. Specifically, we treat the pollster-level outcomes within each state as synthetic copies of that state’s election result and use them to augment the calibration set for conformal PoIP estimation. 
A workflow diagram outlining this localization and enrichment procedure is detained in Figure~\ref{fig: workflow CP PoIP}.

The remaining question is to calculate the conformity scores. For any state $i$ in the target state $s$-localized calibration set $i\in\mathcal{I}_{ca,s}$, let $p_{ijk}^{\rm real}$ denote the hypothetical true polling proportion for candidate $k$ by pollster $j$ in state $i$. This quantity represents the true population-level support rate defined by pollster $j$. It follows that the pollster-specific DoR is given by $d_{ij}^{\rm real} = \log(p_{ij1}^{\rm real})-\log(p_{ij2}^{\rm real})$. This 
leads to a conformity score of the following form: for pollster $j$ in state $i\in\mathcal{I}_{ca,s}$
\[
R_{ij}^{\rm real} = d_{ij}^{\rm real} - \hat d_i.
\]
Consequently, to predict a DoR margin for target state $s$ in the 2024 election, for any potential value $d\in[-1,1]$ of the true DoR margin $d_s^*$, we define $R_s(d) = d - \hat d_i^*$, and the conformal $p$-value function for state $s$ can be computed as
\begin{equation}\label{eq: p-value}
    p_s\big(d\big) = \frac{\sum_{i\in\mathcal{I}_{l_s}}\sum_{j=1}^{n_i} {\bf 1}(R_s(d)< R_{ij}^{\rm real})+1}{\sum_{i\in\mathcal{I}_{l_s}}n_i+1}. 
\end{equation}
We provide the theoretical justification for this $p$-value function in the following theorem, with the proof given in Section S3 of the Supplementary Material.

\begin{theorem}\label{thm: 1}
    Assume the DoR margins $d_s^*$ and $d_{ij}^{\rm real}$ are exchangeable.  Then the p-value function given in (\ref{eq: p-value}) satisfies the following inequality: ${\rm P}[p_s(d_s^*)\le \alpha]\le \alpha$.
\end{theorem}

\begin{remark*}
To interpret the exchangeability assumption between $d_s^*$ and $d_{ij}^{\rm real}$ in Theorem~\ref{thm: 1}, we view $d_s^*$ as the average DoR margin over the {\bf entire voting population} of state $s$ in the 2024 election, and $d_{ij}^{\rm real}$ as the average DoR margin over the {\bf subpopulation sampled by pollster $j$} in state $i$ in the 2016 election, where $i \in \mathcal{I}_{l_s}$ denotes states sharing the same political leaning as state $s$. Then this exchangeability assumption is justified under two key conditions:
(1) the distribution of individual voter preferences in state $s$ during the 2024 election is similar to that in state $i$ during the 2016 election, and
(2) the pollster’s sampling scheme in 2016 is representative of the full voting population in state $i$.
\end{remark*}

In practice, however, the true poll proportions $p_{ij k}^{\rm real}$ are unobservable, as longitudinal follow-up investigations to determine the actual poll rate within individual pollster's sampling frames are typically not viable. Here we propose to approximate $p_{ijk}^{\rm real}$ by the observed poll results $p_{ijk}$ and denote $d_{ij} = \log(p_{ij1})-\log(p_{ij2})$, and this approximation relies on the assumption that the distribution of pollster-level vote shares remains stable between the time of prediction and the final election. 
This leads to an approximate conformity score given as follows: 
\[
R_{ij} = d_{ij} - \hat d_i, \text{ for } i\in \mathcal{I}_{ca,s},
\]
which would retain approximate exchangeability within the localized calibration set. To further account for potential pollster bias, we conduct a sensitivity analysis in Ssection~\ref{sec:sen}. 
A comparison of all versions of conformity scores is presented in Table~\ref{tab: conformity scores}, including the standard, hypothetical and practical conformity scores.

\begin{table}[htbp!]
\centering
{\footnotesize
\begin{tabular}{l|lc|p{6cm}}
\hline\hline
\textbf{Type} & \textbf{Index} & \textbf{Formula} & \textbf{Usage} \\
\hline
\multirow{2}{*}{Standard} 
  & For $i \in \mathcal{I}_{ca}$, 
  & $R_i = R\big((d_i, {\bf x}_i); D_{tr}\big)$ 
  & \multirow{2}{=}{General form in conformal literature (see Supplementary)} \\ 
  & For $s$ and $d \in [-1,1]$, 
  & $R_s(d) = R\big((d, {\bf x}_s^*); D_{tr}\big)$ 
  & \\ 
\hline
\multirow{2}{*}{Hypothetical} 
  & For $i \in \mathcal{I}_{ca,s}$, 
  & $R_{ij}^{\text{real}} = d_{ij}^{\text{real}} - \hat{d}_i$ 
  & \multirow{2}{=}{Hypothetical because $d_{ij}^{\text{real}}$ is unobservable} \\ 
  & For $s$ and $d \in [-1,1]$, 
  & $R_s(d) = d - \hat{d}_i^{*}$ 
  & \\ 
\hline
\multirow{2}{*}{Practical} 
  & For $i \in \mathcal{I}_{ca,s}$, 
  & $R_{ij} = d_{ij} - \hat{d}_i^{*}$ 
  & \multirow{2}{=}{Used in our practice} \\ 
  & For $s$ and $d \in [-1,1]$, 
  & $R_s(d) = d - \hat{d}_i^{*}$ 
  & \\ 
\hline\hline
\end{tabular}}
\caption{Comparison of standard, hypothetical, and practical conformity score formulations with their corresponding usage notes.}
\label{tab: conformity scores}
\end{table}

Replacing $R_{ij}^{\rm real}$ in Equation~(\ref{eq: p-value}) with $R_{ij}$ gives an approximate conformal $p$-value function of the following form: 
\begin{equation}\label{eq: p-value hat}
    \hat p_s\big(d\big) = \frac{\sum_{i\in\mathcal{I}_{l_s}}\sum_{j=1}^{n_i} {\bf 1}(R_s(d)< R_{ij})+1}{\sum_{i\in\mathcal{I}_{l_s}}n_i+1},\;\;\;d\in[-1,1].
\end{equation}
Applying the same steps to develop the inequality~(\ref{poip_5}), in case where $d$ takes the value from the predicted value $\hat d_s^*$, we have the approximation of $\rm PoIP_s$ below:
\begin{eqnarray}
   {\rm PoIP}_s \le  \left\{\begin{array}{cc}1-\hat p_s\big(\hat d_s^*\big),&{\rm if}\; d_s^*<0;\\
\hat p_s\big(\hat d_s^*\big),&{\rm if}\; d_s^*>0.\end{array}\right.\label{eq:conf_p_est}
\end{eqnarray}
The right side of (\ref{eq:conf_p_est}) gives a conformal PoIP estimator that we will use in the rest of paper for our empirical study. 

\begin{figure}[h!tbp]
    \centering
    \includegraphics[width=0.9\linewidth]{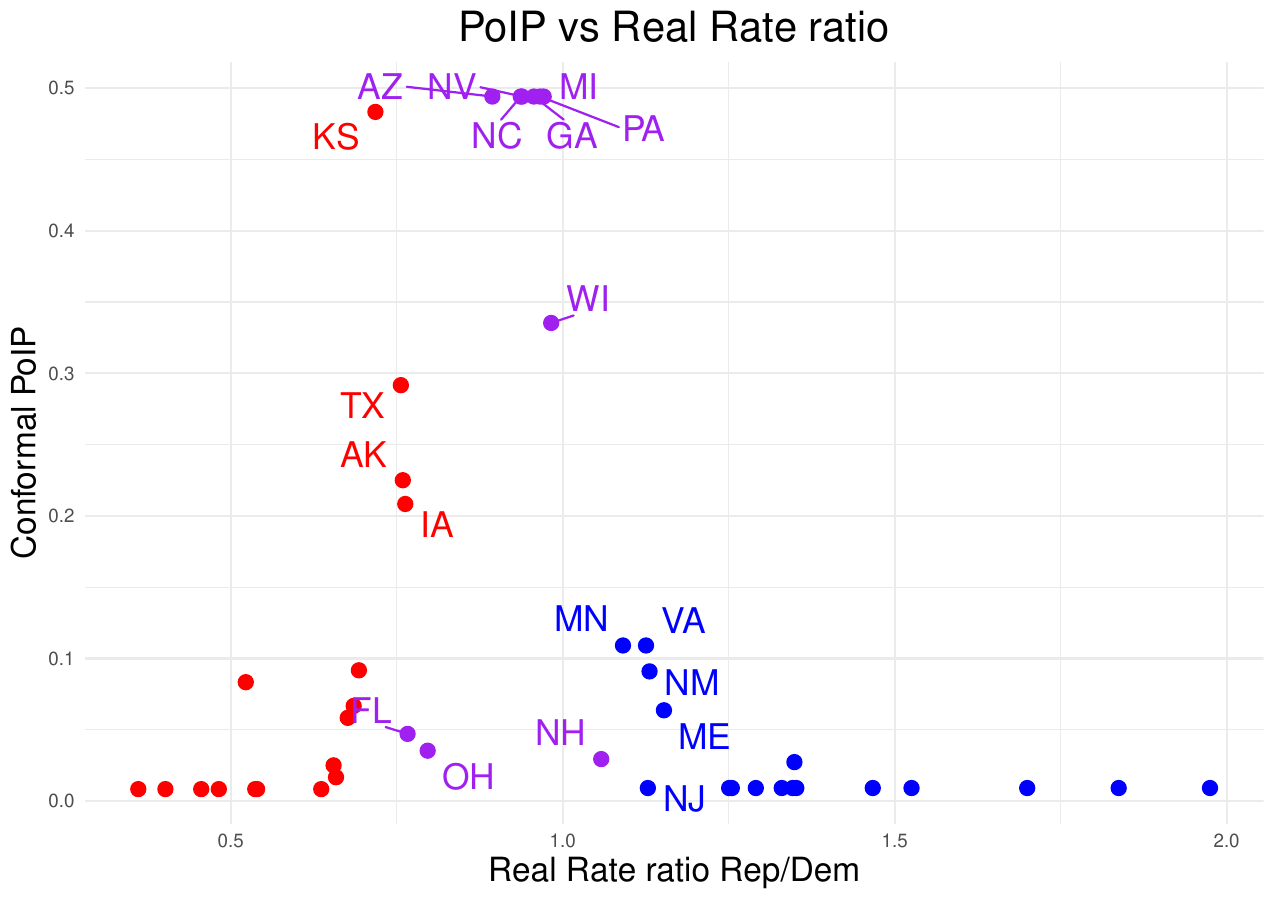}
    \caption{Estimated $\rm PoIP_i$ using conformal prediction method vs odds ratio of vote share for democratic over republic in state $i$ (i.e $\exp(d_i^*)$), for $i$ in red states (red points), blue states (blue points) and swing states (purple points).}
    \label{fig: PoIP CP}
\end{figure}

The conformal PoIP estimates obtained by (\ref{eq:conf_p_est}) are presented in Table~\ref{tab: PoIP boot selected} and displayed in Figure~\ref{fig: PoIP CP}. Table~\ref{tab: PoIP boot selected} shows reasonable magnitudes of estimated PoIP for swing states compared to their bootstrap PoIP counterparts. Figure~\ref{fig: PoIP CP} presents a scatterplot of the estimated conformal PoIP vs the real rate ratio (RRR), $\pi_{i1}/\pi_{i2}$. We can see that, generally, states with RRR close to 1 have larger estimated conformal PoIP (close to $0.5$); those with RRR close to 0 or large RRR have smaller estimated conformal PoIP. Notably, the conformal PoIP estimates for the swing states are generally large. This makes sense because prediction for swing states is known to be hard with many uncontrolled factors, thus their prediction uncertainty should appear larger. Nevertheless, our analysis indicates that three swing states, Florida, Ohio and New Hampshire, have small PoIP,  while several red states, Kansas, Texas, and Alaska have relatively large PoIP. in the past decade or so Florida and Ohio have become substantially Republican-leaning in USPE while Texas, especially in the urban areas of big cities, has received a large number of families moving from blue states. These changes may bias the polling drawn from these states. To explore this issue, we conduct a sensitivity analysis with varing-level of polling bias to examine the robustness of the estimated conformal PoIP.  As shown in Section~\ref{sec:sen},   some of these estimated conformal PoIP values may appear different in the sensitivity analysis.

\subsection{Sensitivity analysis}
\label{sec:sen}

Our analysis has so far ignored potential pollster bias that, unfortunately, cannot be directly measured in practice.  Despite this difficulty, in this section, we conduct a sensitivity analysis to examine to what extent our previous conclusions may be influenced by such bias. 

To overcome the challenge of such unmeasurable bias, we propose an approximation to the bias through its manifestation in the estimation.  Hypothetically, a reasonably unbiased pollster’ sampling strategy would ensure $\mathbb{E}_j\big[d_{ij}^{real}\big]\approx d_i^{real}$; that is, the average of the pollster-specific log-odds differences would be close to the true state-level margin. From this perspective, the pollster bias could be reflected approximately by the difference, $d_{ij}-d_{i}^{real}$, which is termed as the \emph{pollster bias} in this section to navigate our sensitivity analysis.  

Our empirical results reveal that pollster bias varies in both direction and magnitude across election years, and that exhibits distinct patterns across blue, red, and purple states. Detailed visualization of these variations are shown in Figure S2 and Figure S3 in the Supplementary Material.  In particular, for the 2016 election, which is used as the calibration data in our analysis, the winning odds of the Republican candidates were significantly underestimated across nearly all states. This underestimation was most pronounced in red states, and somewhat smaller but still consistent in blue and purple states. In contrast, for the Democratic candidates, their winning odds were underestimated in blue states, but close to the true rate in red and purple states. These empirical results align with existing findings in the election prediction literature \citep{jennings2018election,kennedy2018evaluation,prosser2018twilight}. 

Given the existence of such polling quality issues, it becomes inevitable to assess how such biases may influence our results in the prediction of the 2024 USPE outcomes. Apparently, a sensitivity analysis is deemed necessary and appealing. With minimal information on the distribution of such bias, we propose to use a non-informative prior, that is, assuming that the pollster bias in the 2016 election arises from a uniform distribution: $p_{ijk}-p_{ijk}^{real}\sim{\rm Unif}(a_{l_ik},b_{l_ik})$,
where the limits, $a_{l_ik}$ and $b_{l_ik}$ may be estimated according to the range of observed discrepancies $\{p_{i'jk}-\pi_{i'k}: i'\in \mathcal{I}_{l_i}\}$, with $\mathcal{I}_{l_i}$ denoting the set of states sharing the same political leaning $l_i \in \{\text{blue}, \text{red}, \text{purple}\}$ as state $i$. The estimated limits $a_{l_ik}$ and $b_{l_ik}$ are reported in Table S3 in the Supplementary Material.

\begin{figure}
    \centering
    \includegraphics[width=1\linewidth]{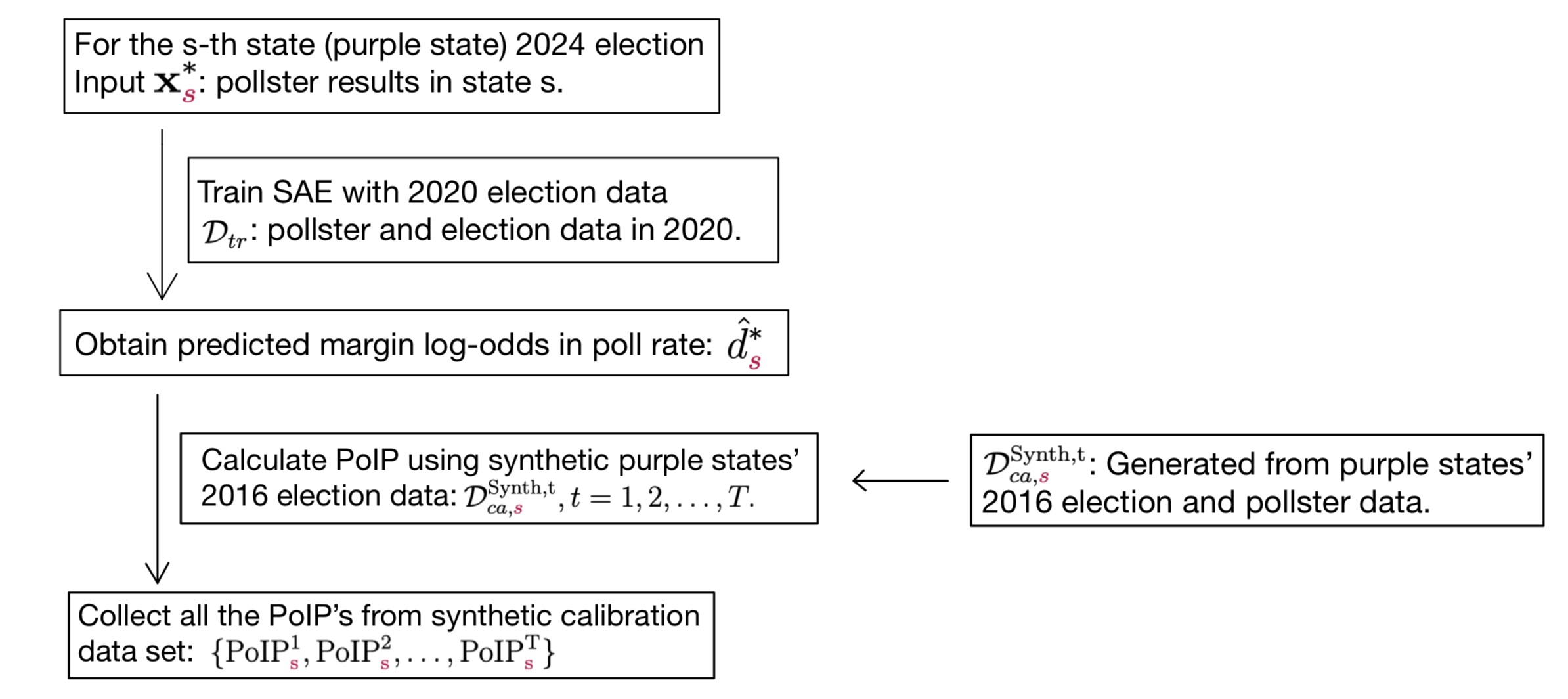}
    \caption{\small{Workflow Diagram for Sensitivity Analysis of Conformal PoIP in a Purple (Swing) State $s \in \mathcal{I}_{\rm purple}$. The Same Procedure Applies to Red and Blue States with Appropriate Calibration Sets.}}
    \label{fig: CP sens diagram}
\end{figure}

Using this uniform distribution for the bias, we generate synthetic poll data $p_{ijk}^{\rm Synth,t} = p_{ijk}-U_{ijk}^{\rm Synth,t}$  with $U_{ijk}^{\rm Synth,t}\sim {\rm Unif}(a_{l_ik},b_{l_ik})$, and then re-compute the conformal PoIP based on the bias-corrected synthetic poll data. The proposed sensitivity analysis workflow is presented in Figure~\ref{fig: CP sens diagram}, including steps implemented for the simulation of the synthetic data and the calculation of conformal PoIP. 

\begin{figure}[h!tbp]
    \centering
    \includegraphics[width=0.99\linewidth]{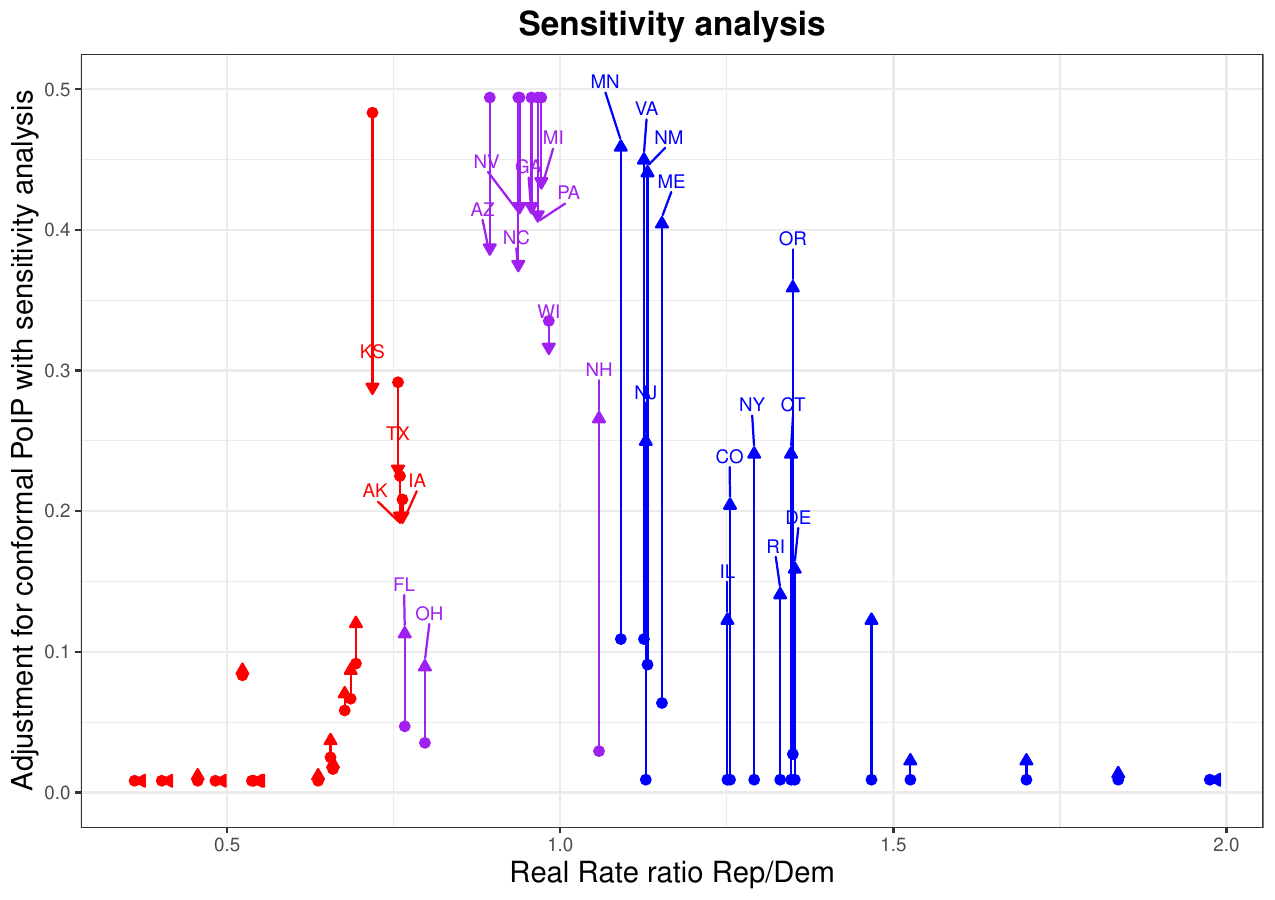}
    \caption{Conformal $\rm PoIP_i$ and Its Adjusted Value from Sensitivity Analysis vs Odds Ratio of Support Rates for Democrats vs Republican in State $i$ (i.e., $\exp(d_i^*)$). Each Point Represents the Originally Estimated PoIP from the Conformal Method, Colored by Political Leaning—Red for Red States, Blue for Blue States, and Purple for Swing States. Arrows Indicate Direction and Magnitude of the Sensitivity Adjustment, with the Arrowheads Marking the Median PoIP Obtained from 200 Simulations under the Bias-adjusted Sensitivity Analysis.}
    \label{fig: sens PoIP}
\end{figure}

Using the workflow, we generate the synthetic data 200 times ($T = 200$) and report the median conformal PoIP values over the 200 simulations in Figure~\ref{fig: sens PoIP}. The results show that the proposed sensitivity analysis provides meaningful adjustments to the conformal PoIP estimates in the previous analysis reported in Table~\ref{tab: PoIP boot selected}. Overall, the adjusted conformal PoIP values become less extreme, moving away from 0 or 0.5.

It is interesting to note that for the red states, the bias adjustments are generally small, except for Kansas (KS), where the adjusted conformal PoIP is substantially lower than the original estimate. This high sensitivity, although the related reasons behind it are unclear, implies a concern about the polling quality in this state, which should be a warning message for the future survey designs in Kansas.  In contrast, for the blue states, the adjusted conformal PoIP values are noticeably higher, suggesting that pollsters may have placed excessively high confidence in these states, and the overconfidence should warrant caution.  Among purple (swing) states that are decisive to recent USPEs, prediction uncertainty tends to decrease after adjusting for the pollster bias. An exception is New Hampshire (NH), where the original conformal PoIP was implausibly small; the bias adjustment corrects this underestimation. In other words, the original conformal PoIP appears to overestimate prediction uncertainty for the swing states. In real-life, such overestimated prediction uncertainty might impact the election campaign.

In summary, the sensitivity analysis results suggest that the pollsters of 2024 USPE may have been overly optimistic in the blue states and unduly pessimistic in the red states, potentially distorting appreciation of prediction confidence. More importantly, the purple states benefit from this noise-injection perturbation in prediction, resulting in a reduced PoIP. Overall, the sensitivity analysis offers valuable insights on variability of confidence for the pollster-based prediction and supports more reliable use of PoIP estimates across states of interest.

Additional results are available in the Supplementary Material, including the 90\% quantile intervals of the conformal PoIP estimates from the 200 sensitivity analysis replicates, as well as the results from the reverse data-split procedure, in which the 2020 election data is used as the calibration set and the 2016 data as the training set (see Section 4.3). The reader are referred to Table S2 and Figure S2 of the Supplementary Material for details.

\section{Conclusion and remarks}
\label{sec:conc}
In this paper, we predict the outcome of the 2024 USPE using pollster data collected one week prior to the election. Our prediction is based on an SAE model trained on data from the 2016 and 2020 elections and polls. The results demonstrate that our prediction achieves 100\% accuracy in predicting the state-level EC winners. To complement the point predictions, we quantify the associated uncertainty using the probability of incorrect prediction (PoIP). We show that the traditional bootstrap methods fail to provide reasonable estimates of the PoIPs. To address the latter issue, we propose a novel conformal prediction-based method, which yields valid and interpretable PoIP estimates. Our findings reveal that states with smaller differences in support rates between the two major parties tend to have higher PoIP values, indicating greater uncertainty and lower prediction confidence.
In addition, we conduct a sensitivity analysis to account for potential pollster bias in the uncertainty quantification. The results show that such bias can meaningfully influence PoIP estimates and should be carefully considered when interpreting the reliability of election predictions. Of course, given that the 2024 USPE results are already known, there is no more uncertainty so far as this election is concerned, but the methods developed in this application are valuable for future practices.

Remarkably, our accurate prediction is achieved using poll data collected just one week prior to the actual election. Such early predictions, when accompanied by principled uncertainty quantification, can have far-reaching post-election implications. For example, reliable predictions can be leveraged for financial gain in the stock market, as investors adjust their portfolios in anticipation of policy changes under different potential administrations. Accurate election prediction can influence campaign strategies, media coverage, and public perception, potentially affecting voter turnout in tight races. Moreover, high-confidence predictions can inform international diplomatic positioning, policy anticipation by industries, and even regulatory timing by government agencies responding to expected political transitions.

There remain directions of further improvements and extension of our proposed methodology. First, additional validation is needed to assess the robustness and generalizability of the SAE model on future election datasets. Second, the SAE framework can be adapted to other types of election prediction tasks by incorporating richer sources of pre-election information (such as geographic or historical voting patterns, demographic trends, or sampling-based pollster data) to enhance prediction accuracy. Third, in the context of our PoIP uncertainty measure, further investigation into pollster-specific bias is warranted to improve the reliability and interpretability of uncertainty quantification.

{\bf Data Availability Statement.} The data used in this paper are publicly available, with the links of the data source given in the Introduction and reference.

\bibliographystyle{agsm}

\bigskip

%
%
%
%
%
%

\end{document}


\def\spacingset#1{\renewcommand{\baselinestretch}%
{#1}\small\normalsize} \spacingset{1}


\if0\blind
{
  \title{{\bf Supplementary Material} for Post-2024 U.S.
  Presidential Election Analysis of Election and Poll Data: Real-life Validation of Prediction via Small Area Estimation and Uncertainty Quantification}
  \author{Zheshi ${\rm Zheng}^{1}$, Yuanyuan ${\rm Li}^{2}$, Peter X. K. ${\rm Song}^{1}$,\\
  and Jiming ${\rm Jiang}^{3}$\\
    University of ${\rm Michigan}^{1}$, Munich ${\rm Re}^{2}$ and\\
    University of California, ${\rm Davis}^{3}$}
  \maketitle
} \fi

\if1\blind
{
  \bigskip
  \bigskip
  \bigskip
  \begin{center}
    {\LARGE\bf Supplementary Materials of Post 2024 U.S. Presidential Election Analysis of Poll Data: Real-life Validation of Prediction using Small Area Estimation}
\end{center}
  \medskip
} \fi

\spacingset{1.8} 
\setcounter{table}{0} 
\renewcommand{\thetable}{S\arabic{table}}
\renewcommand{\thesection}{S\arabic{section}}
\renewcommand{\theequation}{S\arabic{equation}}
\renewcommand{\thefigure}{S\arabic{figure}}


\section{Bootstrap method of estimating ${\rm sd}_{boot}(\hat d_i^*)$.}\label{sec appen: bootstrap}

In this section, we introduce our bootstrap estimate of ${\rm sd}(\hat{d}_i^*)$ as the ${\rm se}_{\rm boot}(\hat{d}_i^*)$. 
Note that the training data (2016 and 2020 election and polling data) under Model I satisfy
\begin{equation}
y_{ia}=\theta_{ia}+e_{ia},\nonumber
\end{equation}
$i=1,\dots,51$, $a=1,\dots,n_{i}$, where $\theta_{ia}=\theta_{ik}=\beta_{0}+\beta_{1}1_{(k=2)}+v_{ik}$, $k=1,2$. Here, $a$ is a combination of $j$ (pollster), $k$ (candidate, $=1,2$), and $t$ (year, $1$ for 2016, $1$ for 2020). The transfer-learning predictor of $d_i^*$ is (see Section 3.1)
\begin{eqnarray}
&&\overline{\log(p_{i\cdot k}^{*})}=\frac{1}{|n_i^*|}\sum_{j=1}^{n_i^*}\log(p_{ijk}^{*}),\nonumber\\
&&\hat d_i^* = \log(\hat\pi_{i1}^*)-\log(\hat\pi_{i2}^*),\,\text{where}\,\, \hat{\pi}_{ik}^*=\exp\{\overline{\log(p_{i\cdot k}^{*})}-\hat{\theta}_{ik}\},\label{poip_7}
\end{eqnarray}
Thus, we need to (i) bootstrap $\hat{\theta}_{ik}$, and (ii) bootstrap $p_{ijk}^{*}$; then, we can use (\ref{poip_7}) to obtain the bootstrapped $\hat d_i^*$.

For (i): We can generate the bootstrapped training data via
\begin{equation}
y_{ia}^{b}=\hat{\theta}_{ia}+e_{ia}^{b}, \, b=1,2,\ldots,B; \, i=1,\dots,51;a=1,\dots,n_{i}\,\nonumber
\end{equation}
where $B=1000$ is the pre-specified number of bootstrap samples, and $e_{ik}^{b}$ are generated independently from $N(0,\hat{\tau}^{2})$ with $\hat{\tau}^{2}$ being the estimate of $\tau^{2}$ based on Model I. We then fit Model I using $y_{ai}^b$ and obtain the bootstrapped $\hat{\theta}_{ia}^b$.

For (ii): Note that each $p_{ijk}^{*}$ is a binomial proportion based on the final survey with survey sample size $n_{ijk}^*$ (known in our data set). The binomial proportion can be bootstrapped by $p_{ijk}^{*b}=Y_{ijk}^{*b}/n_{ijk}^*$, where $Y_{ijk}^{*b}\sim{\rm Binomial}(n_{ijk}^*,p_{ijk}^{*})$.

Then, we obtain the bootstrapped $\hat d_i^*$ by computing
\begin{eqnarray}
&&\overline{\log(p_{i\cdot k}^{*b})}=\frac{1}{|n_i^*|}\sum_{j=1}^{n_i^*}\log(p_{ijk}^{*b}),\nonumber\\
&&\hat d_i^{*b} = \log(\hat\pi_{i1}^{*b}) -\log(\hat\pi_{i2}^{*b}),\,\text{where}\,\, \hat{\pi}_{ik}^{*b}=\exp\{\overline{\log(p_{i\cdot k}^{*b})}-\hat{\theta}_{ik}^{b}\}.\nonumber
\end{eqnarray}
Therefore, we have
\begin{equation}
{\rm se}_{\rm boot}(d_i^*)\approx\sqrt{\frac{1}{B}\sum_{b=1}^{B}\big[\hat{p}^{(b)}-B^{-1}\sum_{b=1}^{B}\hat{d}_i^{*b}\big]}.\nonumber
\end{equation}


\section{Standard split conformal procedure}\label{sec append: std-CP}

Let the observed data be the 2016 and 2020 election and posters' polling data and denote them as $\mathcal{D}_{\rm obs} = \{(\pi_{ik}, p_{ijk})_{j=1,2,\cdots,n_i,k=1,2}: i \in\mathcal{I}_{obs}\}$, where $\mathcal{I}_{obs}$ is the index set for the state $i$ in the observed data set. Denote our prediction target of 2024 election and posters' polling data as $\mathcal{D}_{\rm pred} = \{(\pi_{ik}^*, p_{ijk}^*)_{j=1,2,\cdots,n_i^*,k=1,2}: i \in\mathcal{I}_{pred}\}$, where $\mathcal{I}_{pred}$ is the index set of state $i$ in the 2024 prediction set. Here, $(p_{ijk}^*)_{j=1,2,\ldots, n_i^*,k=1,2}$ are observed at the time of prediction happens and are used to predict the $\pi^*_{ik}$ and the BoR margin $d_s^*$. For any dataset $\mathcal{D}_{\rm index}$, let $\mathcal{I}_{\rm index}$ denote its associated index set. To ensure computational efficiency, we adopt the split conformal procedure \citep{lei2018distribution} in this paper. Specifically, we split the state index $i$ of the observed data into a training set and a calibration set: $\mathcal{D}_{\rm obs} = \mathcal{D}_{\rm tr} \cup \mathcal{D}_{\rm ca}$. The training set $\mathcal{D}_{\rm tr}$ is used to fit the Model I, while the calibration set $\mathcal{D}_{\rm ca}$ is used to calibrate the value of PoIPs.

We calculate a conformity score for each calibration point as $R_i = R\big((d_i,{\bf x}_i);\mathcal{D}_{\rm tr}\big)$ for $i\in\mathcal{I}_{\rm ca}$, here ${\bf x}_i = (p_{ijk})_{j=1,2,\ldots,n_i,k=1,2}$ is the polling data of the $i$-th state. For any prediction index $s\in\mathcal{I}_{\rm pred}$ and any candidate value $d_s^* = d$, we denote $R_s(d) = R\big((d,{\bf x}^*_s);\mathcal{D}_{\rm tr}\big)$ obtained by plugging $(d,{\bf x}_s^*)$ into the conformity score function, and ${\bf x}^*_s = (p_{sjk})_{j=1,2,\ldots,n^*_s,k=1,2}$ is the 2024 polling data of the $s$-th state. 
Under the assumption that the calibration data $\mathcal{D}_{\rm ca}$ and prediction data $\mathcal{D}_{\rm pred}$ are exchangeable (i.e., they are drawn from the same distribution and their joint distribution is invariant under permutations of the sample order), the $R_s(d)$ and $\{R_i:i\in\mathcal{I}_{ca}\}$ are exchangeable when $d$ is close to the true DoR margin unobserved value $d^*_s$. Under the exchangeability assumption, the conformal $p$-value function takes the form: 
\begin{equation}
p_s(d) = p(d; {\bf x}^*_s) = \frac{\sum_{i\in\mathcal{I}_{ca}} {\bf 1}(R_s(d)< R_i)+1}{|\mathcal{I}_{ca} |+1}, \label{eq:conf_pvalue1}
\end{equation}
where ${\bf 1}(\cdot)$ is the indicator function. This conformal p-value function has well-established theoretical properties and can be used to construct valid prediction intervals for the target prediction quantity \citep{xie2022homeostasis}.

\section{Proof of Theorem~1}\label{sec appen: CP thm}
For notation simplicity, write $R_{s,1}^{\rm real} = R_s(d_s^*) = d_s^*-\hat d_s^*$ with $n_s = 1$, and let $\mathcal{I}_{l_s}^* = \mathcal{I}_{l_s}\cup\{s\}$. By the assumption that $d_s^*$ and $d_{ij}^{real}$ are exchangeable, we have $R_s^*$ and $R_{ij}^{\rm real}$ are exchangeable. Then
\begin{align*}
    P&\big[p_s(d_s^*)\le \alpha\big] = P\left[\frac{\sum_{i\in\mathcal{I}_{l_s}}\sum_{j=1}^{n_i} {\bf 1}(R_s(d_s^*)< R_{ij}^{\rm real})+1}{\sum_{i\in\mathcal{I}_{l_s}}n_i+1}\leq \alpha\right]\\    &=P\left[\frac{\sum_{i\in\mathcal{I}_{l_s}^*}\sum_{j=1}^{n_i} {\bf 1}(R_{s,1}^{real}\le  R_{ij}^{\rm real})}{\sum_{i\in\mathcal{I}_{l_s}^*}n_i}\leq \alpha\right]=\mathbb{E} {\bf 1}\left\{\frac{\sum_{i\in\mathcal{I}_{l_s}^*}\sum_{j=1}^{n_i} {\bf 1}(R_{s,1}^{real}\le  R_{ij}^{\rm real})}{\sum_{i\in\mathcal{I}_{l_s}^*}n_i}\leq \alpha\right\}\\
    & = \frac{1}{\sum_{i\in\mathcal{I}_{l_s}^*}n_i} \mathbb{E}\sum_{t\in\mathcal{I}_{l_s}^*}\sum_{j'=1}^{n_t}{\bf 1}\left\{\frac{1}{\sum_{i\in\mathcal{I}_{l_s}^*}n_i}\sum_{i\in\mathcal{I}_{l_s}^*}\sum_{j=1}^{n_i}{\bf 1}(R_{tj'}^{real}\le  R_{ij}^{\rm real})\le \alpha\right\}
\end{align*}

Order the statistics $\{R_{ij}^{real}\}_{i\in\mathcal{I}_{l_s},j=1,2,\ldots,n_i}$ by $R_{(1)}<R_{(2)}<\ldots<R_{(N)}$, and denote $N_i$ the number of $R_j$'s that equal to $R_{(i)}$, and define $N_0=0$, then $\sum_{i=0}^N N_i = \sum_{i\in\mathcal{I}_{l_s}^*}n_i$. 

Then for $R_{t'j'} = R_{(t)}$, we have $\sum_{i\in\mathcal{I}_{l_s}^*}\sum_{j=1}^{n_i}{\bf 1}(R_{t'j'}^{\rm real}\ge R_{ij}^{\rm real}) = \sum_{i=0}^i N_t $. For any $\alpha\in (0,1)$, there exists $T\in\{0,1,\ldots,N-1\}$ such that $\sum_{i=0}^T N_i\le  \sum_{i\in\mathcal{I}_{l_s}^*}n_i\alpha < \sum_{i=0}^{T+1} N_i$. Thus
$$
\frac{1}{\sum_{i\in\mathcal{I}_{l_s}^*}n_i} \mathbb{E}\sum_{t\in\mathcal{I}_{l_s}^*}\sum_{j'=1}^{n_t}  {\bf 1}\left\{\frac{1}{\sum_{i\in\mathcal{I}_{l_s}^*}n_i}\sum_{i\in\mathcal{I}_{l_s}^*}\sum_{j=1}^{n_i}{\bf 1}(R_{tj'}^{real}\le  R_{ij}^{\rm real})\le \alpha\right\} = \frac{\sum_{i=1}^T N_i}{\sum_{i\in\mathcal{I}_{l_s}^*}n_i}\le\alpha
$$ 
Therefore proves the theorem.

\section{Supplementing tables and figures}
\subsection{Detailed SAE model prediction results}
The detailed prediction results from the SAE models are presented in Table~\ref{tab:predictions}. Our findings show that two of the proposed SAE models, Model I and Model III, achieve smaller prediction errors compared to the PoP method, with Models I and III producing nearly identical outcomes. Moreover, in terms of predicting the winner in each state, the SAE models achieve perfect accuracy, correctly identifying the winning candidate in all 44 states. In contrast, the PoP method mis-predicts three key battleground states: Michigan, Pennsylvania, and Wisconsin, resulting in an incorrect prediction of the overall national winner.
\begin{table}[htbp]
    \centering
    \renewcommand{\arraystretch}{0.8}
    \captionsetup{skip=2pt}
    \resizebox{0.85\textwidth}{!}{
    \begin{tabular}{l|cccc|cccc|c}
        \toprule
        State & Actual Votes-D & SAE-M1-D & SAE-M3-D & PoP-D & Actual Vote-R& SAE-M1-R & SAE-M3-R& PoP-R & EC Votes \\
        \midrule
        AK        & 41.4 & 44.2 & 44.7 & 43.4 & 54.5 & 58.2 & 58.8 & 52.0 & 3  \\
        AZ       & 46.7 & 48.2 & 48.2 & 46.9 & 52.2 & 52.6 & 52.6 & 48.4 & 11 \\
        AR      & 33.6 & 40.7 & 41.3 & 40.0 & 64.2 & 61.2 & 62.0 & 55.0 & 6  \\
        CA    & 58.5 & 64.7 & 65.0 & 59.9 & 38.3 & 35.3 & 35.3 & 34.2 & 54 \\
        CO      & 54.1 & 57.8 & 58.0 & 54.3 & 43.1 & 45.2 & 45.3 & 42.0 & 10 \\
        CT   & 56.4 & 55.1 & 56.9 & 53.0 & 41.9 & 42.9 & 44.2 & 37.0 & 7  \\
        DE      & 56.6 & 56.6 & 58.9 & 54.8 & 41.9 & 41.3 & 42.7 & 36.5 & 3  \\
        FL       & 43.0 & 45.1 & 45.1 & 44.8 & 56.1 & 56.8 & 56.8 & 51.2 & 30 \\
        GA       & 48.5 & 48.9 & 48.9 & 47.4 & 50.7 & 52.2 & 52.2 & 48.5 & 16 \\
        IL      & 54.4 & 63.3 & 64.3 & 58.4 & 43.5 & 44.2 & 44.6 & 41.6 & 19 \\
        IN       & 39.6 & 41.4 & 41.3 & 41.0 & 58.6 & 64.3 & 64.6 & 55.7 & 11 \\
        IA          & 42.5 & 44.5 & 44.4 & 45.2 & 55.7 & 58.0 & 57.9 & 50.8 & 6  \\
        KS        & 41.0 & 43.0 & 42.9 & 43.2 & 57.2 & 54.7 & 54.6 & 48.2 & 6  \\
        ME         & 52.4 & 50.6 & 51.2 & 50.3 & 45.5 & 49.3 & 49.9 & 41.5 & 4  \\
        MD      & 62.6 & 64.4 & 65.3 & 60.5 & 34.1 & 35.5 & 35.7 & 33.4 & 10 \\
        MA & 61.2 & 63.7 & 65.3 & 60.8 & 36.0 & 37.2 & 37.9 & 33.0 & 11 \\
        MI      & 48.3 & 49.0 & 48.9 & 48.2 & 49.7 & 52.3 & 52.3 & 46.6 & 15 \\
        MN     & 50.9 & 51.3 & 51.1 & 50.1 & 46.7 & 48.9 & 48.8 & 44.2 & 10 \\
        MO      & 40.1 & 40.8 & 40.6 & 41.6 & 58.5 & 62.8 & 62.7 & 54.4 & 10 \\
        MT       & 38.5 & 37.1 & 36.9 & 38.2 & 58.4 & 65.2 & 65.1 & 56.4 & 4  \\
        NE      & 39.1 & 39.1 & 40.3 & 39.0 & 59.6 & 64.3 & 66.3 & 55.4 & 5  \\
        NV        & 47.5 & 50.0 & 50.1 & 47.5 & 50.6 & 51.1 & 51.3 & 48.0 & 6  \\
        NH & 50.7 & 54.0 & 54.2 & 52.4 & 47.9 & 49.4 & 49.6 & 44.4 & 4  \\
        NJ    & 52.0 & 57.3 & 57.9 & 54.7 & 46.1 & 44.2 & 44.6 & 39.9 & 14 \\
        NM    & 51.9 & 53.5 & 54.3 & 49.9 & 45.9 & 46.0 & 46.6 & 43.0 & 5  \\
        NY      & 55.9 & 60.2 & 61.1 & 56.7 & 43.3 & 43.2 & 43.6 & 39.6 & 28 \\
        NC& 47.7 & 47.7 & 47.6 & 47.3 & 50.9 & 52.5 & 52.5 & 48.3 & 16 \\
        ND  & 30.5 & 35.4 & 36.4 & 36.0 & 67.0 & 64.3 & 66.5 & 54.5 & 3  \\
        OH          & 43.9 & 44.3 & 44.4 & 44.3 & 55.1 & 59.2 & 59.3 & 52.1 & 17 \\
        OK      & 31.9 & 33.0 & 33.3 & 34.0 & 66.2 & 72.8 & 73.9 & 62.6 & 7  \\
        OR        & 55.3 & 55.3 & 55.9 & 53.0 & 41.0 & 45.2 & 45.6 & 41.0 & 8  \\
        PA  & 48.7 & 48.3 & 48.2 & 47.8 & 50.4 & 52.1 & 52.0 & 47.6 & 19 \\
        RI  & 55.5 & 55.8 & 60.2 & 54.0 & 41.8 & 41.2 & 44.2 & 35.8 & 4  \\
        SC& 40.4 & 42.6 & 42.4 & 42.1 & 58.2 & 59.9 & 59.8 & 53.8 & 9  \\
        SD  & 34.2 & 32.7 & 32.8 & 34.8 & 63.4 & 75.1 & 75.9 & 60.5 & 3  \\
        TN     & 34.5 & 34.7 & 34.9 & 36.7 & 64.2 & 72.4 & 73.5 & 58.9 & 11 \\
        TX         & 42.5 & 46.0 & 45.9 & 44.6 & 56.1 & 55.9 & 55.9 & 51.4 & 40 \\
        UT          & 37.8 & 35.7 & 36.0 & 36.0 & 59.4 & 69.0 & 69.5 & 57.8 & 6  \\
        VT       & 63.8 & 64.2 & 66.1 & 63.0 & 32.3 & 36.9 & 37.9 & 31.0 & 3  \\
        VA      & 51.8 & 52.4 & 52.6 & 50.1 & 46.1 & 47.7 & 47.8 & 43.8 & 13 \\
        WA    & 57.2 & 59.0 & 59.2 & 56.4 & 39.0 & 40.1 & 40.2 & 37.4 & 12 \\
        WV & 28.1 & 31.8 & 31.6 & 34.0 & 70.0 & 73.6 & 73.8 & 61.0 & 4  \\
        WI     & 48.8 & 47.3 & 47.1 & 48.3 & 49.7 & 53.4 & 53.3 & 46.9 & 10 \\
        WY       & 25.8 & 25.9 & 26.3 & 27.5 & 71.6 & 80.0 & 81.8 & 66.2 & 3  \\
        \bottomrule
    \end{tabular}
    }
    \caption{Predicted and actual vote percentages for 2024 US presidential election (-D and -R are for democratic party and republican party, respectively). SAE-M1 represent model 1, SAE-M3 represent model 3. PoP is the "poll of polls" method.}
    \label{tab:predictions}
\end{table}

\subsection{Detailed PoIP estimation using bootstrap and conformal methods}

The PoIP estimates obtained using both the bootstrap and conformal methods are presented in Table~\ref{tab: PoIP full results}. Our results show that the bootstrap method consistently produces small PoIP values across all states, even including battleground states, thereby substantially underestimating the true uncertainty in the predictions. In contrast, the proposed conformal method yields significantly higher PoIP estimates for battleground states, better reflecting the inherent uncertainty in closely contested races. Those PoIPs are further corrected by the sensitivity analysis, which takes the pollster bias into account.

\begin{table}[htbp]
    \centering
    \renewcommand{\arraystretch}{0.8}
    \captionsetup{skip=2pt}
    \resizebox{0.6\textwidth}{!}{
    \begin{tabular}{{l|rrrrrrr}}
\toprule
state & realrate OR & SAE OR & PoP OR & bootstrap PoIP & conf. PoIP & sensitivity intervals\\
\midrule\centering
WY & 0.3609 & 0.3244 & 0.4147 & $<0.0001$ & 0.0083 & [0.0083, 0.0083] \\
WV & 0.4017 & 0.4325 & 0.5574 & $<0.0001$ & 0.0083 & [0.0083, 0.0167] \\
ND & 0.4556 & 0.5495 & 0.6606 & $<0.0001$ & 0.0083 & [0.0083, 0.0333] \\
OK & 0.4822 & 0.4530 & 0.5437 & $<0.0001$ & 0.0083 & [0.0083, 0.0167] \\
AR & 0.5227 & 0.6653 & 0.7273 & $<0.0001$ & 0.0833 & [0.0583, 0.1333] \\
TN & 0.5370 & 0.4787 & 0.6228 & $<0.0001$ & 0.0083 & [0.0083, 0.0250] \\
SD & 0.5398 & 0.4348 & 0.5744 & $<0.0001$ & 0.0083 & [0.0083, 0.0167] \\
UT & 0.6364 & 0.5181 & 0.6234 & $<0.0001$ & 0.0083 & [0.0083, 0.0333] \\
NE & 0.6550 & 0.6079 & 0.7043 & $<0.0001$ & 0.0250 & [0.0167, 0.0667] \\
MT & 0.6587 & 0.5694 & 0.6779 & $<0.0001$ & 0.0167 & [0.0083, 0.0417] \\
IN & 0.6763 & 0.6396 & 0.7367 & $<0.0001$ & 0.0583 & [0.0417, 0.1088] \\
MO & 0.6852 & 0.6499 & 0.7646 & $<0.0001$ & 0.0667 & [0.0583, 0.1254] \\
SC & 0.6931 & 0.7106 & 0.7815 & $<0.0001$ & 0.0917 & [0.0750, 0.1750] \\
KS & 0.7180 & 0.7877 & 0.8963 & $<0.0001$ & 0.4833 & [0.2246, 0.3500] \\
TX & 0.7563 & 0.8215 & 0.8667 & $<0.0001$ & 0.2917 & [0.1750, 0.2833] \\
AK & 0.7593 & 0.7595 & 0.8279 & $<0.0001$ & 0.2250 & [0.1417, 0.2500] \\
IA & 0.7630 & 0.7682 & 0.8905 & $<0.0001$ & 0.2083 & [0.1417, 0.2417] \\
FL & 0.7664 & 0.7945 & 0.8759 & $<0.0001$ & 0.0471 & [0.0824, 0.1647] \\
OH & 0.7967 & 0.7492 & 0.8504 & $<0.0001$ & 0.0353 & [0.0588, 0.1353] \\
AZ & 0.8941 & 0.9175 & 0.9690 & $<0.0001$ & 0.4941 & [0.3235, 0.4529] \\
NC & 0.9369 & 0.9087 & 0.9796 & $<0.0001$ & 0.4941 & [0.3174, 0.4412] \\
NV & 0.9387 & 0.9775 & 0.9914 & 0.0377 & 0.4941 & [0.3529, 0.4768] \\
GA & 0.9566 & 0.9379 & 0.9772 & $<0.0001$ & 0.4941 & [0.3471, 0.4768] \\
PA & 0.9661 & 0.9270 & 1.0048 & $<0.0001$ & 0.4941 & [0.3471, 0.4765] \\
MI & 0.9714 & 0.9362 & 1.0351 & $<0.0001$ & 0.4941 & [0.3765, 0.4941] \\
WI & 0.9827 & 0.8849 & 1.0302 & $<0.0001$ & 0.3353 & [0.2647, 0.3944] \\
NH & 1.0581 & 1.0934 & 1.1792 & $<0.0001$ & 0.0294 & [0.2294, 0.3235] \\
MN & 1.0908 & 1.0483 & 1.1342 & 0.0020 & 0.1091 & [0.3905, 0.4909] \\
VA & 1.1255 & 1.0999 & 1.1445 & $<0.0001$ & 0.1091 & [0.3727, 0.4909] \\
NJ & 1.1283 & 1.2962 & 1.3720 & $<0.0001$ & 0.0091 & [0.1909, 0.3091] \\
NM & 1.1309 & 1.1614 & 1.1615 & $<0.0001$ & 0.0909 & [0.3636, 0.4909] \\
ME & 1.1525 & 1.0265 & 1.2108 & 0.1133 & 0.0636 & [0.3364, 0.4909] \\
IL & 1.2507 & 1.4316 & 1.4038 & $<0.0001$ & 0.0091 & [0.0818, 0.1909] \\
CO & 1.2548 & 1.2783 & 1.2936 & $<0.0001$ & 0.0091 & [0.1455, 0.2727] \\
NY & 1.2909 & 1.3937 & 1.4322 & $<0.0001$ & 0.0091 & [0.1818, 0.3091] \\
RI & 1.3300 & 1.3524 & 1.5105 & $<0.0001$ & 0.0091 & [0.1000, 0.2000] \\
CT & 1.3464 & 1.2837 & 1.4324 & $<0.0001$ & 0.0091 & [0.1818, 0.3091] \\
OR & 1.3490 & 1.2238 & 1.2927 & $<0.0001$ & 0.0273 & [0.2905, 0.4455] \\
DE & 1.3519 & 1.3701 & 1.5034 & $<0.0001$ & 0.0091 & [0.1182, 0.2364] \\
WA & 1.4671 & 1.4712 & 1.5094 & $<0.0001$ & 0.0091 & [0.0818, 0.1909] \\
CA & 1.5254 & 1.8324 & 1.7485 & $<0.0001$ & 0.0091 & [0.0091, 0.0545] \\
MA & 1.6996 & 1.7150 & 1.8410 & $<0.0001$ & 0.0091 & [0.0091, 0.0545] \\
MD & 1.8374 & 1.8151 & 1.8089 & $<0.0001$ & 0.0091 & [0.0091, 0.0364] \\
VT & 1.9749 & 1.7383 & 2.0323 & $<0.0001$ & 0.0091 & [0.0091, 0.0182] \\
\bottomrule
    \end{tabular}
    }
    \caption{Complete table of the estimated PoIP from the bootstrap and conformal approach (bootstrap PoIP and conf. PoIP), with the 90\% interval from sensitivity analysis with $200$ repetitions. Reported quantities include the odds ratios of the real-rate (realrate OR), the SAE-estimated odds ratios (SAE OR), the PoP odds ratios (PoP OR). States are ordered by the odds ratios of the real-rate (realrate OR.).}
    \label{tab: PoIP full results}
\end{table}

\subsection{Reverse analysis using 2020 election data as calibration data set}

\begin{figure}[htbp]
    \centering
    \includegraphics[width=0.9\linewidth]{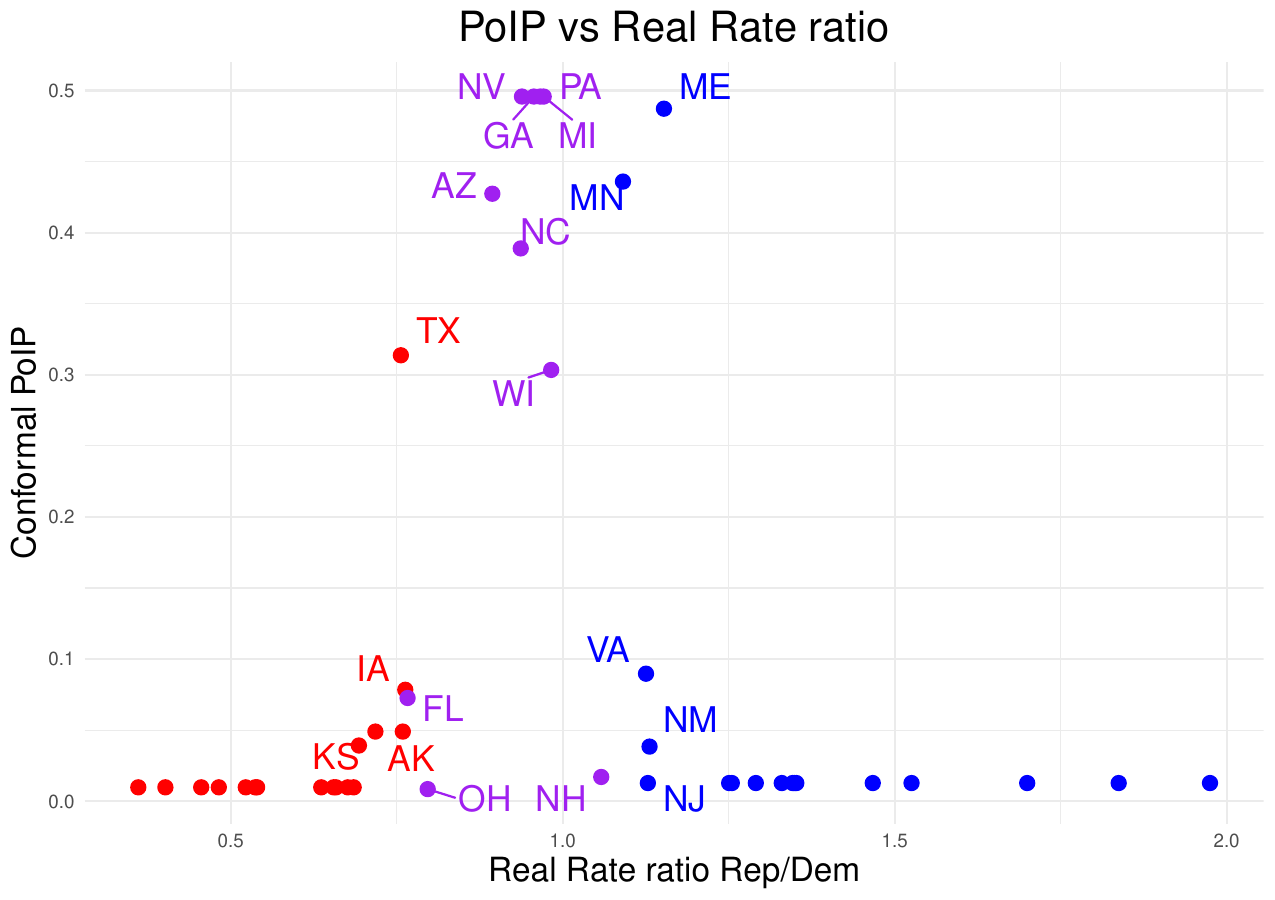}
    \caption{Estimated $\rm PoIP_i$ using conformal prediction method using 2020 U.S. election data as calibration vs odds ratio of vote share for democratic over republic in state $i$ (i.e $\exp(d_i^*)$), for $i$ in red states (red points), blue states (blue points) and swing states (purple points).}
    \label{fig: CP PoIP 2020 calib}
\end{figure}

\begin{figure}
    \centering
    \includegraphics[width=0.9\linewidth]{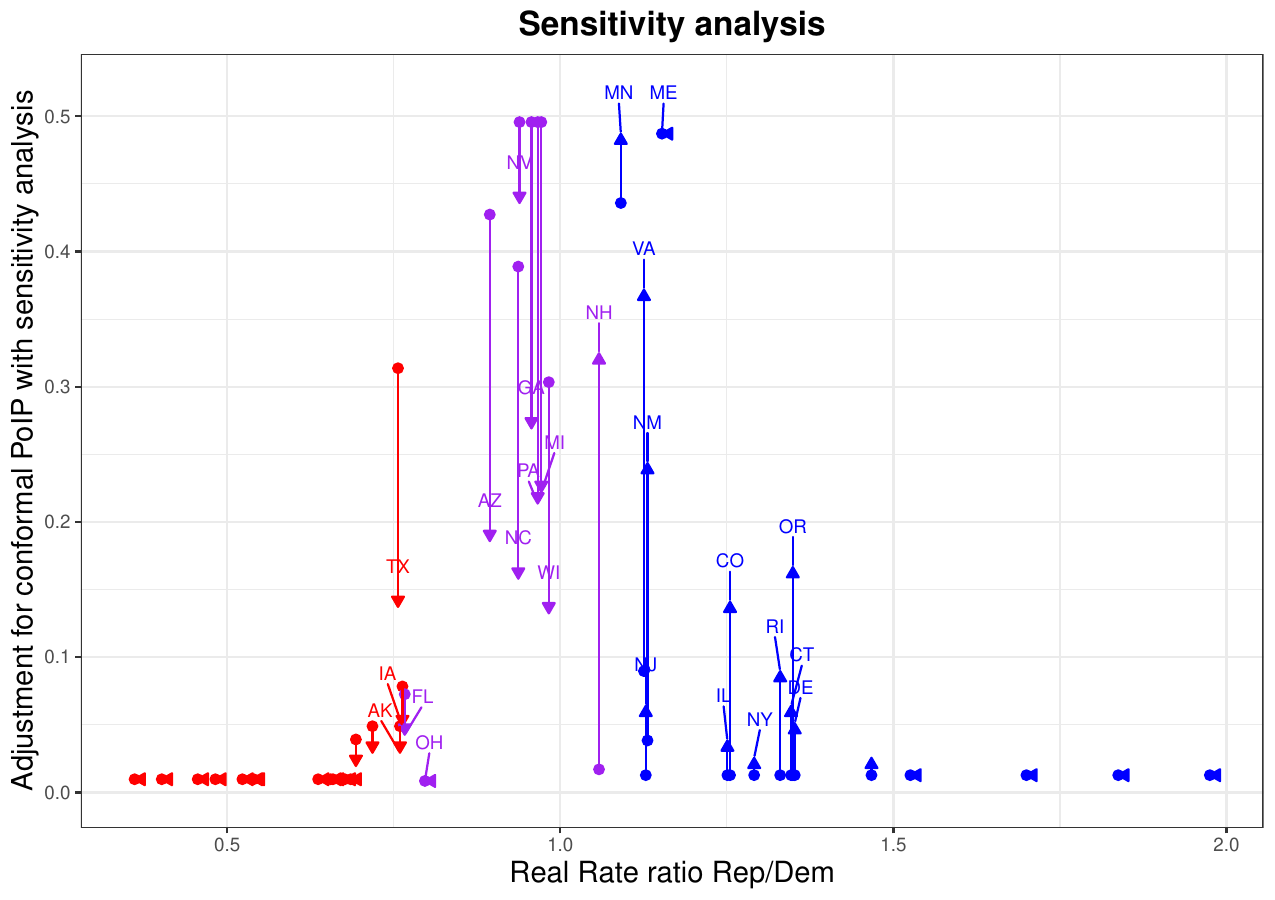}
    \caption{Absolute real reat difference vs PoIP under case 1. Blue dots are PoIP calculated from pollsters rates and intervals are from 90\% quantiles of the 200 repetitions from the sensitivity analysis.}
    \label{fig:sensitivity analysis 2020 calib}
\end{figure}

We conduct an additional reverse analysis to estimate PoIP using the conformal prediction method introduced in the main text. In this analysis, we use the 2016 election data as the training set and the 2020 pollster results as the calibration set. The resulting PoIP estimations are presented in Figure~\ref{fig: CP PoIP 2020 calib}. The results closely align with those obtained using the 2016 data as the calibration set, suggesting that the year-based data-splitting strategy is robust to the choice of election year. For the sensitivity analysis, results are given in ~\ref{fig:sensitivity analysis 2020 calib}, the conclusions are the same as using the 2016 data to do the calibration.

\subsection{Upper and lower bounds of the bias in the sensitivity analysis}
In this section, we analyze the pollster bias and present the estimated upper and lower bound used in the sensitivity analysis.

\begin{figure}[htbp]
    \centering
    \includegraphics[width=0.95\linewidth]{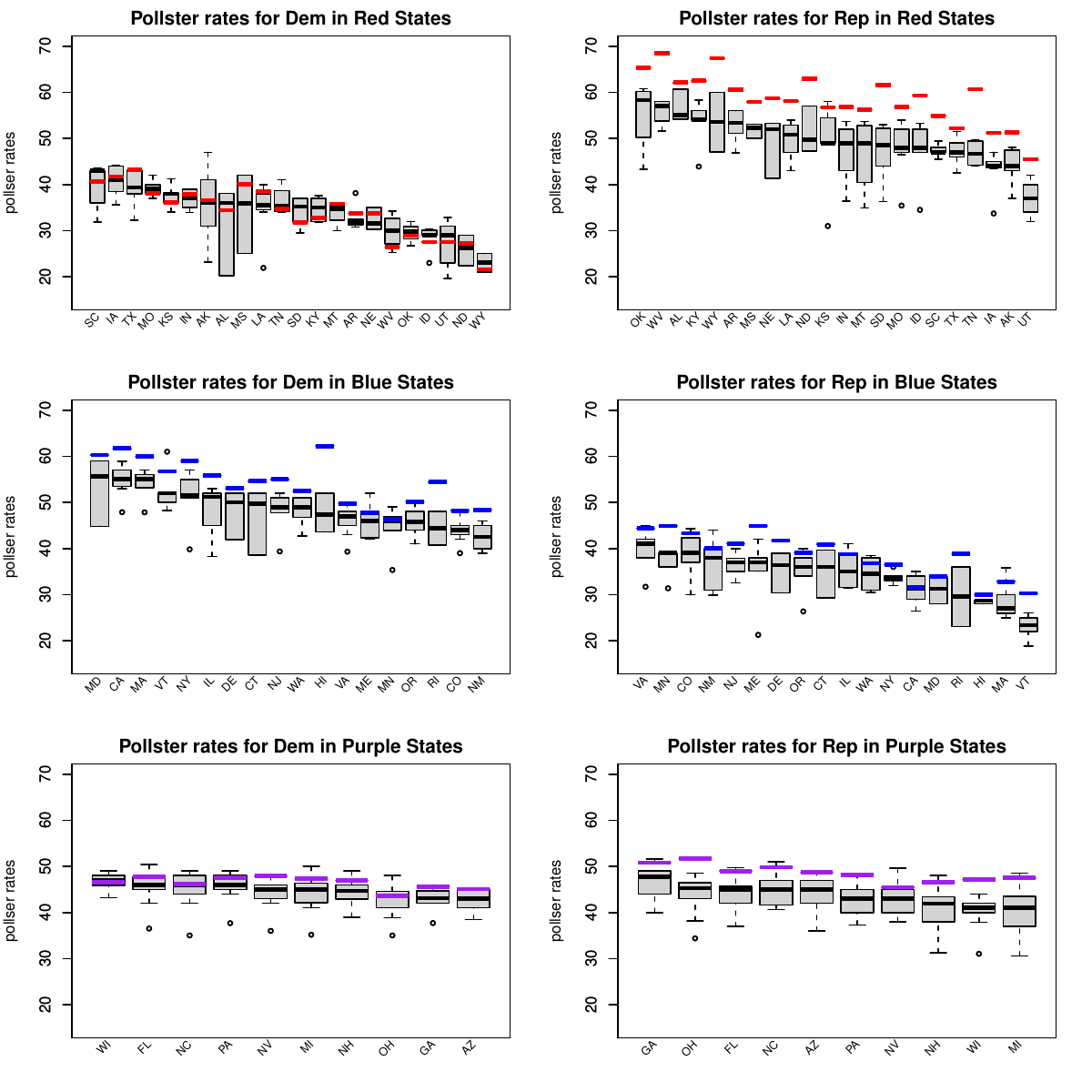}
    \caption{Pollster bias in U.S. states during the 2016 presidential election. Each boxplot represents the distribution of pollster-reported vote share differences within a given state. Colored horizontal lines indicate the actual vote share margin in that state: red for Republican-leaning states, blue for Democratic-leaning states, and purple for swing states.}
    \label{fig:pollster bias 16}
\end{figure}

\begin{figure}[htbp]
    \centering
    \includegraphics[width=0.95\linewidth]{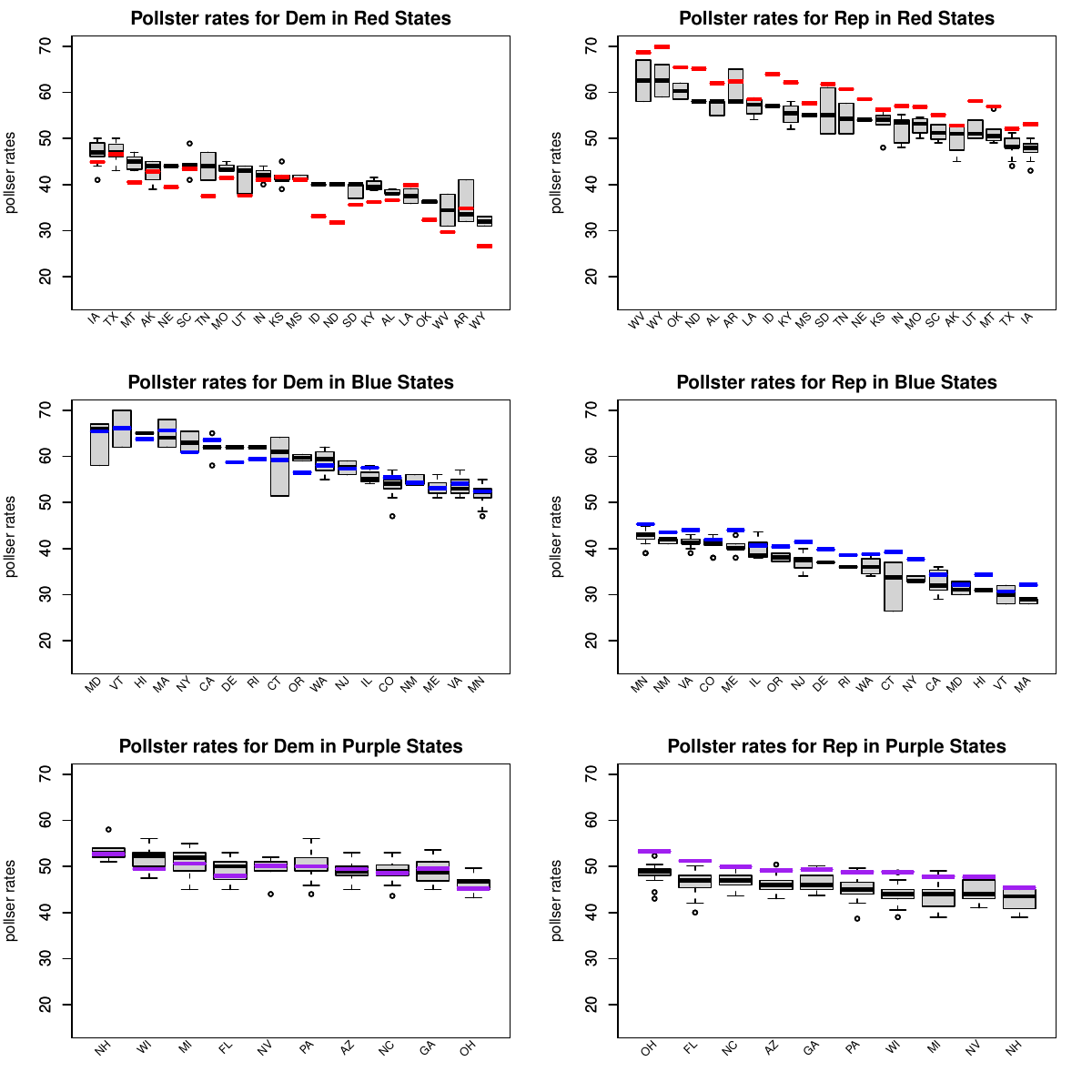}
    \caption{Pollster bias in U.S. states during the 2020 presidential election. Each boxplot represents the distribution of pollster-reported vote share differences within a given state. Colored horizontal lines indicate the actual vote share margin in that state: red for Republican-leaning states, blue for Democratic-leaning states, and purple for swing states.}
    \label{fig:pollster bias 20}
\end{figure}

We begin by visualizing the distribution of vote shares reported by pollsters across states using boxplots, shown in Figures~\ref{fig:pollster bias 16} and~\ref{fig:pollster bias 20}. In 2016, Republican candidates were significantly underestimated across nearly all states. This underestimation was most pronounced in red states, and somewhat smaller but still consistent in blue and purple states. For Democratic candidates, they were underestimated in blue states, while estimates in red and purple states were relatively close to the real rate.
In 2020, the bias pattern shifted. Pollsters tended to overestimate support for Democratic candidates and underestimate support for Republican candidates. The bias against Republican candidates remained large across nearly all states. For Democratic candidates, the overestimation bias was more substantial in red states, while it was relatively smaller in blue and purple states.

We present the estimated upper and lower bounds of pollster bias used to generate synthetic data in the sensitivity analysis in Table~\ref{tab: sens a_kj and b_jk}.

\begin{table}[htbp]
\centering
\begin{tabular}{rrrrlc}
\hline\hline
$b_{l_i1}$ & $a_{l_i1}$ & $b_{l_i2}$ & $a_{l_i2}$ & color $l_i$ & Year \\
\hline
19.17 &  -4.30 & 23.62 & -4.00 & Blue   & 2016 \\
16.58 & -10.40 & 25.70 & -1.30 & Red    & 2016 \\
12.12 &  -4.40 & 17.30 & -4.11 & Purple & 2016 \\
\hline
 8.40 &  -4.90 & 12.80 & -3.00 & Blue   & 2020 \\
 4.00 &  -9.50 & 10.90 & -2.60 & Red    & 2020 \\
 6.10 &  -6.60 & 11.20 & -1.30 & Purple & 2020 \\
\hline\hline
\end{tabular}
\caption{Estimated $a_{l_ik}$ and $b_{l_ik}$ for $l_i\in\{\rm Blue, Red, Purple\}$, $k=1$ for Democratic and $k=2$ for Republican, in both year 2016 and year 2020.}
\label{tab: sens a_kj and b_jk}
\end{table}
